\documentclass[rmp,aps,nofootinbib,superscriptaddress]{revtex4-1}

\pdfoutput=1

\usepackage{color}
\usepackage{times}
\usepackage{graphicx}
\usepackage{fancyhdr}
\usepackage{float}
\usepackage{url}

\def\be{\begin{equation}}
\def\ee{\end{equation}}
\def\bi{\begin{itemize}}
\def\ei{\end{itemize}}
\def\ben{\begin{enumerate}}
\def\een{\end{enumerate}}

\begin{document}

\title{Multimessenger astronomy with gravitational waves and high-energy neutrinos}

\author{Shin'ichiro Ando}
\affiliation{Gravitation AstroParticle Physics Amsterdam Institute (GRAPPA), The Netherlands}
\author{Bruny Baret}
\affiliation{AstroParticule et Cosmologie (APC), CNRS: UMR7164-IN2P3-Observatoire de Paris-Universit\'e Denis Diderot-Paris VII-CEA: DSM/IRFU, France}
\author{Imre Bartos}
\affiliation{Department of Physics, Columbia University, New York, NY 10027, USA}
\author{Boutayeb Bouhou}
\affiliation{AstroParticule et Cosmologie (APC), CNRS: UMR7164-IN2P3-Observatoire de Paris-Universit\'e Denis Diderot-Paris VII-CEA: DSM/IRFU, France}
\author{Eric Chassande-Mottin}
\affiliation{AstroParticule et Cosmologie (APC), CNRS: UMR7164-IN2P3-Observatoire de Paris-Universit\'e Denis Diderot-Paris VII-CEA: DSM/IRFU, France}
\author{Alessandra Corsi}
\affiliation{Physics Department, George Washington University, Washington, D.C. 20052, USA}
\affiliation{LIGO Laboratory, California Institute of Technology,
Pasadena, CA 91125, USA}
\author{Irene Di Palma}
\affiliation{Albert-Einstein-Institut, Max-Planck-Institut f\"ur Gravitationsphysik, D-30167 Hannover, Germany}
\author{Alexander Dietz}
\affiliation{Department of Physics and Astronomy of the University of Mississippi, Mississippi 38677-1848, USA}
\author{Corinne Donzaud}
\affiliation{AstroParticule et Cosmologie (APC), CNRS: UMR7164-IN2P3-Observatoire de Paris-Universit\'e Denis Diderot-Paris VII-CEA: DSM/IRFU, France} 
\affiliation{Universit\'e  Paris-sud, Orsay, F-91405, France}
\author{David Eichler}
\affiliation{Department of Physics, Ben Gurion University, Beer-Sheva 84105, Israel}
\author{Chad Finley}
\affiliation{Oskar Klein Centre \& Dept. of Physics, Stockholm University, SE-10691 Stockholm, Sweden}
\author{Dafne Guetta}
\affiliation{Department of Physics and Optical Engineering, ORT Braude, P.O. Box 78, Karmiel, Israel}
\affiliation{INAF-Observatory of Rome, Via Frascati 33, Monteporzio Catone, Italy}
\author{Francis Halzen}
\affiliation{Department of Physics, University of Wisconsin, Madison, WI 53706,USA}
\author{Gareth Jones}
\affiliation{School of Physics and Astronomy,Cardiff University, Cardiff CF24 3AA, United Kingdom}
\author{Shivaraj Kandhasamy}
\affiliation{University of Minnesota, Minneapolis, MN 55455, USA}
\author{Kei Kotake}
\affiliation{Division of Theoretical Astronomy, National Astronomical Observatory of Japan, 2-21-1,Osawa, Mitaka, Tokyo, 181-8588, Japan}
\affiliation{Department of applied physics, Fukuoka University,  Jonan, Fukuoka,  814-0189, Japan}
\author{Antoine Kouchner}
\affiliation{AstroParticule et Cosmologie (APC), CNRS: UMR7164-IN2P3-Observatoire de Paris-Universit\'e Denis Diderot-Paris VII-CEA: DSM/IRFU, France}
\author{Vuk Mandic}
\affiliation{University of Minnesota, Minneapolis, MN 55455, USA}
\author{Szabolcs M\'arka}
\affiliation{Department of Physics, Columbia University, New York, NY 10027, USA}
\author{Zsuzsa M\'arka}
\affiliation{Department of Physics, Columbia University, New York, NY 10027, USA}
\author{Luciano Moscoso}
\email{deceased}
\affiliation{AstroParticule et Cosmologie (APC), CNRS: UMR7164-IN2P3-Observatoire de Paris-Universit\'e Denis Diderot-Paris VII-CEA: DSM/IRFU, France}
\author{Maria Alessandra Papa}
\affiliation{Albert-Einstein-Institut, Max-Planck-Institut f\"ur Gravitationsphysik, D-30167 Hannover, Germany}
\author{Tsvi Piran}
\affiliation{Racah Institute of Physics, Hebrew University of Jerusalem, Jerusalem 91904, Israel}
\author{Thierry Pradier}
\affiliation{Universit\'e de Strasbourg \& Institut Pluridisciplinaire Hubert Curien,Strasbourg, France}
\author{Gustavo E. Romero}
\affiliation{Instituto Argentino de Radioastronomia (IAR, CCT La Plata, CONICET), C.C. No. 5, 1894, Villa Elisa, Buenos Aires, Argentina}
\affiliation{FCAyG, Observatorio de La Plata, Paseo del Bosque s/n, CP 1900 La Plata, Argentina.}
\author{Patrick Sutton}
\affiliation{School of Physics and Astronomy,Cardiff University, Cardiff CF24 3AA, United Kingdom}
\author{Eric Thrane}
\affiliation{University of Minnesota, Minneapolis, MN 55455, USA}
\author{V\'eronique Van Elewyck}
\email{Corresponding author's electronic address: elewyck@apc.univ-paris7.fr}
\affiliation{AstroParticule et Cosmologie (APC), CNRS: UMR7164-IN2P3-Observatoire de Paris-Universit\'e Denis Diderot-Paris VII-CEA: DSM/IRFU, France}
\author{Eli Waxman}
\affiliation{Department of Particle Physics \& Astrophysics, The Weizmann Institute of Science, Rehovot 76100, Israel }

\vspace*{1.5cm}

\begin{abstract}
Many of the astrophysical sources and violent phenomena observed in our Universe are potential emitters of gravitational waves and high-energy cosmic radiation, including photons, hadrons, and presumably also neutrinos. Both gravitational waves (GW) and high-energy neutrinos (HEN) are cosmic messengers that may escape much denser media than photons. They travel unaffected over cosmological distances, carrying information from the inner regions of the astrophysical engines from which they are emitted (and from which photons and charged cosmic rays cannot reach us). For the same reasons, such messengers could also reveal new, hidden sources that have not been observed by conventional photon-based astronomy.

Coincident observation of GWs and HENs may thus play a critical role in multimessenger astronomy.  This is particularly true at the present time owing to the advent of a new generation of dedicated detectors: the neutrino telescopes IceCube at the South Pole and ANTARES in the Mediterranean Sea, as well as the GW interferometers Virgo in Italy and LIGO in the United States. Starting from 2007, several periods of concomitant data taking involving these detectors have been conducted. More joint datasets are expected with the next generation of advanced detectors that are to be operational by 2015, with other detectors, such as KAGRA in Japan, joining in the future. Combining information from these independent detectors can provide original ways of constraining the physical processes driving the sources, and also help confirm the astrophysical origin of a GW or HEN signal in case of coincident observation. 

Given the complexity of the instruments, a successful joint analysis of this combined GW+HEN observational dataset will be possible only if the expertise and knowledge of the data is shared between the two communities. This review aims at providing an overview of both theoretical and experimental state of the art and perspectives for GW+HEN multimessenger astronomy. 
\end{abstract}

\maketitle

\tableofcontents

\section{Introduction}  
High-energy multimessenger astronomy has entered an exciting era with the development and operation of new detectors offering unprecedented opportunities to observe cosmic radiation in the Universe in all its variety. 
Gamma-ray astronomy has been an illustrative example of the synergy between particle physics and astronomy. Soon after the discovery of the neutral pion, \citet{1952PThPh...8..571H} suggested that the interaction of cosmic rays with the neutral gas in the Galactic Plane should produce a diffuse, extended gamma-ray emission. Almost simultaneously, \citet{Hutchinson52} calculated the contribution to such an emission from relativistic bremsstrahlung. Prospects for gamma-ray astronomy were set up shortly after by~\citet{Morrison58}. The first gamma-ray sources were discovered during the late 1960s and 1970s, but even until the 1990s they were difficult to identify, calling for multi-wavelength observational efforts (see~\citet{2004ASSL..304.....C} for an overview of the subject). During the past decade, this approach has revealed itself to be fruitful in the identification of several types of sources from MeV to TeV energy scales. 

The mechanisms that produce the high-energy radiation have, however, remained elusive, requiring the development of multi-messenger techniques and programs that would explore all components of the cosmic radiation (see collective references in~\citet{Paredes2007}). In the study of transient sources, which involve compact objects and ultra-violent phenomena (such as gamma-ray bursts and magnetars), multimessenger techniques are in fact  the only approaches that might lead to a full understanding of the underlying processes. In this context, high-energy ($\gg$GeV) neutrinos (HENs) and gravitational waves (GWs) could play an important role.  These messengers share interesting astronomical properties: HENs can escape from much denser, hence deeper, environments than photons, and GWs propagate virtually freely in any region of space. Moreover, and contrary to high-energy photons (which can be absorbed by intervening photon backgrounds) and charged cosmic rays (which are deflected by ambient magnetic fields),  both GW and HEN propagate at the speed of light through magnetic fields and matter without being altered. Therefore, they are expected to provide important information about the processes taking place in  astrophysical engines, and could even reveal the existence of sources opaque to hadrons and photons, sources that would thus far have remained undetected. While neither HENs nor GWs have been directly observed to date, it is widely believed that a first detection could plausibly occur in the near future; see e.g.~\citet{Becker:2007sv} and~\citet{0264-9381-28-11-114013} for reviews on these subjects. This colloquium is dedicated to the prospects for astronomy using these two cosmic messengers.

Many astrophysical sources, the majority of which originate from cataclysmic events, are expected to produce both GWs and HENs. While GWs are linked to the dynamics of the bulk motion of the source progenitor, HENs trace the interactions of accelerated protons (and possibly heavier nuclei) with matter and radiation in and around the source.  
An overview of the most plausible sources of HENs and GWs is presented in Section~\ref{sec:science} of this article, along with relevant references. It includes transient sources such as  extra-galactic 
gamma-ray bursts (GRBs), for which popular progenitor models involve either the collapse of a highly-rotating massive star or the merger of a binary system of compact objects (neutron star/neutron star or black hole/neutron star); both of these  scenarios are expected to be associated with the emission of GWs.  The presence of accelerated hadrons in the jets emitted by the source  would ensure the subsequent production of HENs. Magnetars, though less powerful sources, are closer (galactic) and more frequently occurring; they are also considered as possible GW+HEN emitters. Observation-based phenomenological arguments bounding the time delay between the GW and HEN emission in the sources are presented in Section \ref{sec:timewindow}.

The current efforts carried out for the detection of GWs and HENs are described in Section~\ref{sec:detection}. Concerning the detection of neutrinos, huge ($\sim$km$^3$) volumes of target material need to be monitored to compensate for the feeble signal expected from plausible astrophysical sources. Current neutrino telescopes are in-water or in-ice Cherenkov detectors which rely on the construction of  3D arrays of photomultiplier tubes.  IceCube\footnote{\citet{icecubeGen}; see also \url{http://icecube.wisc.edu}}  
is a km$^3$-scale detector located at the geographic South Pole, while \textsc{Antares}\footnote{\citet{antaresGen}; see also \url{http://antares.in2p3.fr}}, 
with an instrumented volume $\sim 0.02$ km$^3$, is deployed undersea, 40 km off the French coast and serves as a prototype for a future km$^3$-scale detector in the Mediterranean. The 
combination of the two detectors provides full coverage of the sky and partial redundancy. The direct detection of GWs is performed through the operation of large ($\sim$km scale) laser interferometers. Several GW observatories have been operating recently: the two LIGO\footnote{\citet{Abbott:2007kv}; see also \url{http://www.ligo.org}}  detectors in the USA (one in Livingston, 
Louisiana, another in Hanford, Washington), Virgo\footnote{\citet{accadia12:_virgo}; see also \url{http://www.virgo.infn.it}}  near Pisa (Italy) and GEO\footnote{\citet{Grote:2010zz}; see also \url{http://geo600.aei.mpg.de}} near Hanover (Germany), 
collectively form a network of detectors that allows for the localisation of astrophysical sources. 
Both HEN and GW detectors have been developing multimessenger strategies that involve other cosmic probes, in particular electromagnetic radiation in a wide range of wavelength bands. These  are typically based either on the use of external triggers (such as GRB events) or on follow-up programs; more detail is given in Sections~\ref{MMGW} and~\ref{MMHEN}.

Neither GWs nor cosmic HENs have been individually detected so far. The detection of coincident GW and HEN events would hence be a landmark observation and help confirm the astrophysical origin of both signals. Coincident searches are also a way to enhance the sensitivity of the joint detection channel by exploiting the correlation between HEN and GW significances, taking advantage of the fact that the two types of detectors have uncorrelated backgrounds. Since a joint analysis requires a consistent signal to be observed in both instruments in space as well as time, there is a significant background suppression relative to each individual analysis, hence an increased discovery potential. Preliminary investigations of the feasibility of such searches have been performed by  \citet{ligo_icecube} and \citet{Pradier:2008uj} and indicate that, even if the constituent observatories provide several triggers a day, the false alarm rate for the combined detector network can be maintained at a very low level, e.g. 1/(600 yr) for some realistic parameters. 

A  major challenge for the analysis lies in the combined optimisation of the selection criteria for the different detection techniques. Section~\ref{sec:gwhen} starts with laying the basics of the data analysis procedures used in each experiment, including the performance of the detectors, and  concentrating on the important aspects connected to GW+HEN searches such as the accuracy of the source sky position reconstruction. Different options for a combined GW+HEN analysis are then presented. Section~\ref{subsec:trig} describes a method for a HEN-triggered GW search: in this case, the search for GW signals is performed only in parts of the sky defined by neutrino candidate events, and within a time window defined by the observational and phenomenological considerations discussed in Section~\ref{sec:timewindow}. 
The outcome of such an analysis, performed recently with data from the construction phase of ANTARES and from the initial LIGO/VIRGO detectors, is also presented.
Alternatively, comprehensive searches for space-time coincidences between independent lists of neutrino and GW events can also be performed, as illustrated in Section~\ref{subsec:parallel} through an example baseline search that could be performed with IceCube and LIGO/Virgo. In this case, time-coincident signals are tested for correlation using a combined GW+HEN likelihood skymap, as well as additional information on the individual significance of the HEN and GW candidates (such as their spatial correlation with large-scale matter distribution). This second, more symmetric and comprehensive option requires the existence of two independent analysis chains scanning the whole phase space in search of interesting events. 
Both Sections~\ref{subsec:trig} and ~\ref{subsec:parallel}  illustrate how these searches can be used to infer limits on the population of astrophysical GW+HEN sources, while the Conclusion presents general perspectives for the astrophysical reach of GW+HEN searches.

The joint search activities described in this paper are performed in the framework of a dedicated  GW+HEN working group involving collaborators from all the previously mentioned experiments. The data-exchange policies are regulated by specific bilateral Memoranda of Understanding. 


\section{The science case for multimessenger GW+HEN searches}

\subsection{Potential emitters of GW and HEN}
\label{sec:science}

\subsubsection{Galactic sources: soft gamma-ray repeaters}
 
Soft gamma-ray repeaters (SGRs) are X-ray pulsars which have quiescent soft (2-10 keV) periodic X-ray emissions with periods ranging from 
5 to 10 s. They exhibit repetitive erratic bursting episodes lasting a few hours each and composed of numerous very short ($\sim$ ms) pulses. Every once in a while they emit a giant flare in which a short ($<0.5$ sec) spike of harder radiation is observed; such flares can reach peak luminosities of $\sim 10^{47}$ erg/s, in X-rays and $\gamma$-rays.
A handful of SGR sources are known, most of them in the Milky Way and one in the Large Magellanic Cloud. Their detected population has been increasing in the last years, thanks to more sensitive instruments and better monitoring\footnote{See e.g.~\citet{Hurley:1998ks}; \citet{1999ApJ...519L.143H}; \citet{2000ApJ...531..407C}; \citet{2003ApJ...585..948K}; \citet{2005Natur.434.1107P}; \citet{2008A&ARv..15..225M}; \citet{2009ApJ...698L..82A}; \citet{2010MmSAI..81..432H}; \citet{2010ApJ...718..331G}; \citet{2010ApJ...710.1335K}; \citet{2010ApJ...711L...1V}.}. Three of the known SGRs have had hard spectrum ($\sim$ MeV energy) giant flares: one with a luminosity of $10^{47}$ erg/s, the two others being two orders of magnitude weaker. 

The {\it magnetar} model describes these objects as a neutron star with an enormous magnetic field $B \gtrsim 10^{15}$ G which can be subject to star-quakes that are thought to fracture the rigid crust, causing outbursts~\cite{Duncan:1992hi,Thompson:1995gw,Thompson:1996pe} . The giant flares result from the formation
 and dissipation of strong localized currents due to magnetic field rearrangements that are associated with the quakes, and liberate a high flux of 
X- and $\gamma$-rays. Sudden changes in the large magnetic fields would accelerate protons or nuclei that produce neutral and charged pions in interactions with thermal radiation. These hadrons would subsequently decay into TeV or even PeV energy $\gamma$-rays and neutrinos~\cite{sgrnu_1, sgrnu_2}, making flares from SGRs potential sources of HENs. 
An alternative model involving a large scale rearrangement of the magnetic field has also been proposed by~\citet{Eichler:2002fc}, which allows for huge energy releases, and detectable HEN fluxes from Galactic magnetars even for relatively  small HEN efficiencies.


 
 During the crustal disruption, a fraction of the initial magnetic energy
is annihilated and released as photons, and the stored elastic energy is
also converted into shear vibrations. SGR flares may excite to some
extent the fundamental or f-modes of the star, which radiate GW with
damping times of $\sim 200$ ms, as described e.g. by~\textcite{1983ApJS...53...73L},~\citet{sgrgw} and~\citet{2004PhRvD..70h4009G}. These timescales are shorter
than other relevant ones, except for the Alfv\'en-wave crossing time of the
star, to which they are comparable.
If much of the flare energy goes into exciting the f-modes, they might
emit GW energy exceeding the emitted EM energy.

Detailed predictions about the GW amplitude are difficult to obtain. An
upper limit of $\sim10^{49}$ erg on the maximum \textit{total} energy release in
an SGR giant flare can be derived from one of the most optimistic models
(\citet{Ioka:2001}, Fig. 3) of a giant flare  associated with a global
reconfiguration of the internal magnetic field \cite{Eichler:2002fc}. Similarly,
\citet{Corsi:2011zi} estimated a maximum total energy release of
$\sim10^{48}$-$10^{49}$ ergs in a fraction of the parameter space, within the model
originally proposed by \citet{Ioka:2001}. To date, the best LIGO f-mode limit
is $1.4\times 10^{47}$\,erg (at 1090 Hz and a nominal distance of 1 kpc) for SGR
0501+451 \cite{Abadie:2010wx}. This upper limit
probes below the most optimistic total energy estimates, but most likely
only a small fraction of the total available energy actually goes into
GWs. Recent works by~\citet{2011PhRvD..83h1302K},~\citet{2011MNRAS.418..659L},~\citet{Ciolfi:2012en},~\citet{ Lasky:2012ju} and~\citet{ 2012PhRvD..85b4030Z} suggest
that the fraction of flare energy that goes into exciting the f-mode is
very small ($E_{GW} \ll 10^{45}$ erg for magnetic fields smaller than 10$^{16}$ G),
making the prospect for a detection of GWs from SGR f-modes with the
advanced LIGO and Virgo unlikely. However, this question may be open to
further investigations; see e.g.~\citet{2011PhRvD..83h1302K},~\citet{2011MNRAS.418..659L},~\citet{Lasky:2012ju} and~\citet{ 2012PhRvD..85b4030Z}.

\subsubsection{Gamma-ray bursts}
\label{sec:grb}

Gamma-ray Bursts (GRBs) are detected as an intense and short-lived flash of gamma-rays with energies ranging from tens of keVs to tens of GeVs. The morphology of their light curves is highly variable and typically exhibits millisecond variability, suggesting very compact sources and relativistic expansion. GRBs are divided into two classes depending on the duration of their prompt gamma-ray emission, which appears to be correlated with the hardness of their spectra and are believed to arise from different progenitors:  short-hard bursts last less than 1 --  2 seconds  (depending on the observing detector) while  long-soft bursts can last up to dozens of minutes.

The BATSE detector, launched in 1991 on board the Compton Gamma-Ray Observatory, was the first mission to accumulate observations on more than a thousand GRBs, establishing the isotropy of their sky distribution and characterizing their light curve and broken power-law spectra \cite{BATSE4B}. The detection of X-ray and optical counterparts pertaining to the afterglow phase of several GRBs, triggered by the first observation of an X-ray transient emission from GRB970228  by the BeppoSAX satellite~\cite{1997Natur.387..783C}, subsequently confirmed their extragalactic origin by allowing more accurate localization of the source and redshift determination.
Currently operating GRB missions include {\it Swift}~\cite{2004ApJ...611.1005G}, hosting a wide-field hard X-ray (15 keV - 350 keV) burst alert telescope (BAT) coupled to softer X-ray, ultraviolet and optical telescopes and the GBM on the Fermi Gamma-Ray Space Telescope~\cite{Atwood:2009ez} which focuses on the high-energy (15 keV - 300 GeV) emission from GRBs.

\begin{figure}[!t]
\vspace*{-0.5cm}
\includegraphics[width=0.45\linewidth, height=6cm]{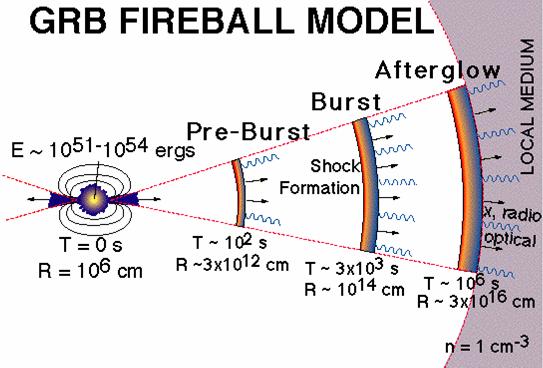}
\hfill
\includegraphics[width=0.40\linewidth, height=6cm]{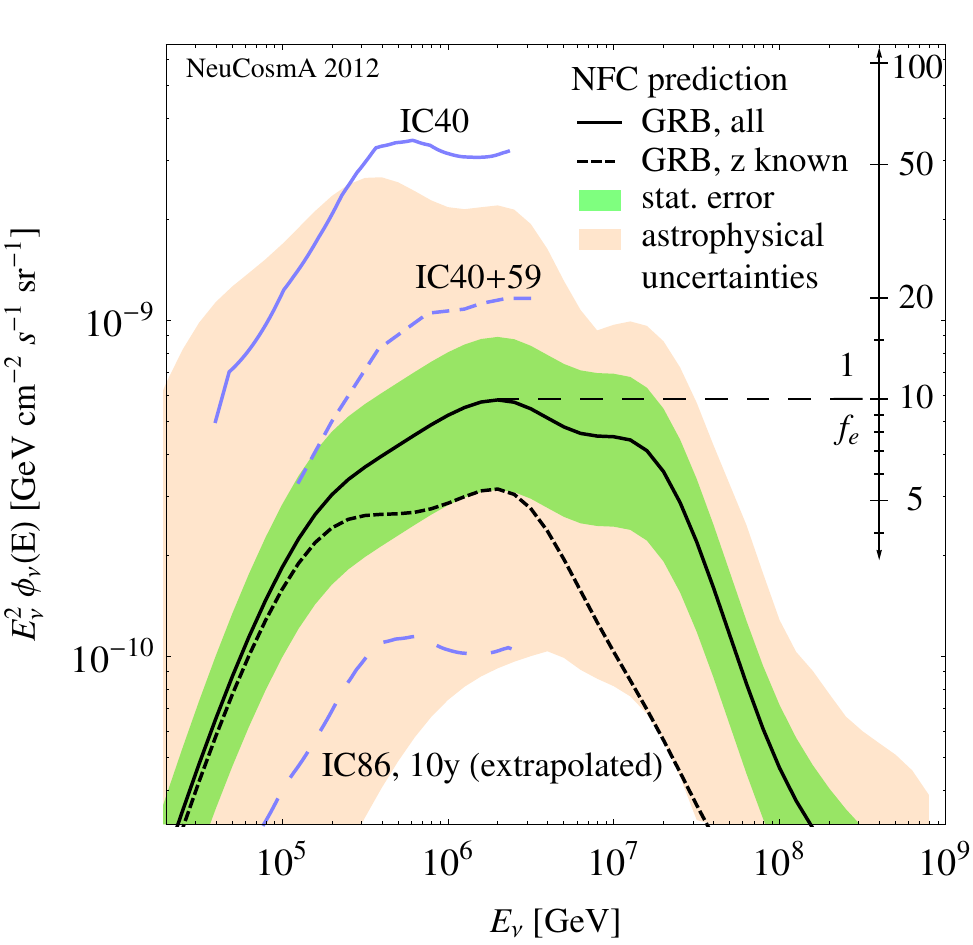}
\caption{\textbf{Left:} Schematic depiction of the fireball mechanism, with the characteristic time and distance scales associated with the different phases. The prompt (burst) phase is due to internal shocks in the relativistically expanding fireball, producing strong gamma-ray and X-ray emission.  The afterglow arises from the cooling fireball and its interaction with the surrounding medium; it is associated with X-ray, optical and radio emission. From www.swift.ac.uk . \textbf{Right:} Current limits set by IceCube in its 40-string and 59-string configurations (IC40 and IC40+59) on the quasi-diffuse GRB neutrino flux, together with predictions based on recent numerical calculations taking into account uncertainties on the astrophysical parameters and those due to the limited statistics of bursts. Also shown is the extrapolated limit that one can expect from 10 years of operation with the full IceCube detector (IC86). From~\citet{Hummer:2011ms}.}
\label{fig:grb}
\end{figure}

In the standard picture, the mechanism responsible for
the enormous, super-Eddington energy release ($\sim 10^{50} - 10^{52}$ ergs) in the prompt emission and in the afterglow 
is the dissipation (via internal shocks, magnetic reconnection and external shocks)
of bulk kinetic or Poynting flux  into highly relativistic particles; see e.g. \citet{Meszaros:1993tv} and the review by \citet{Piran:2004ba}. The particles are accelerated to a non-thermal energy distribution via the Fermi mechanism in a relativistically expanding fireball ejected by the GRB central engine, as sketched in Fig.~\ref{fig:grb} (left panel). The accelerated electrons (and positrons) in the intense magnetic field emit non-thermal photons via synchrotron radiation and inverse Compton scattering.
The plasma parameters inferred from observations to characterize GRB baryonic fireballs are such that proton acceleration to energies exceeding $10^{20}$eV is likely to be possible in these sources. Moreover, the time averaged energy output of GRBs in photons is comparable to the proton energy production rate required to produce the UHECR flux\footnote{The latter statement has been criticized by several authors recently, arguing that the GRB energy production rate is too small to account for the flux of ultrahigh-energy cosmic rays (UHECRs); see e.g. \cite{Eichler:2010ky}. However, the validity of this criticism has been challenged by \citet{Waxman:2010fj}.}. Therefore, the canonical baryonic fireball also suggests that  GRBs are a prime candidate source for the UHECR, observed at energies $E\sim 10^{18} - 10^{20}$ eV ~\cite{1993ApJ...418..386L,Waxman:1995vg, Vietri:1995hs}. 

In a baryonic outflow, the internal or external shocks accelerate protons that  interact  with the gamma-rays and/or other protons inside the fireball, producing charged pions and kaons that  subsequently decay into HENs ( $\pi^{\pm}, K^{\pm} \rightarrow \mu^{\pm} + \nu_{\mu}/\overline{\nu}_{\mu} \rightarrow e^\pm + \nu_e/\overline{\nu_e} + \nu_\mu/\overline{\nu}_{\mu})$\footnote{Relevant references on these mechanisms include \citet{1994ApJS...90..877E}, \citet{Paczynski:1994uv}, \citet{WBfireball}, \citet{1998PhRvD..58l3005R}, \citet{2000PhRvD..62i3015A}, \citet{2001PhRvL..87q1102M}, \citet{2001PhRvL..87q1102M}, \citet{2003PhRvL..90t1103G}, \citet{2003PhRvL..90x1103R}, \citet{2003PhRvD..68h3001R}, \citet{2003PhRvL..91g1102D}, \citet{guetta04:_neutr_batse}, \citet{2005PhRvL..95f1103A}, \citet{2006PhRvL..97e1101M}, \citet{2006ApJ...651L...5M}.}. %
Such neutrinos are emitted in spatial and temporal coincidence with the GRB prompt electromagnetic signal; their energy is typically in the range $\sim $ TeV to PeV. Neutrinos with higher (up to $\sim 10^{10}$ GeV) energy can also be emitted at the beginning of the afterglow phase, when the outflow is decelerated by external shocks with ambient material and the accelerated protons undergo interactions with the matter outside of the jet~\cite{WBafterglow2000}. An alternative mechanism for neutrino production in fireballs suggested by~\citet{levinson2003} involves neutral particles that are picked up by the stream when they acquire a charge, such as a decaying neutron or, further downstream, a neutral atom that is ionized. Such a particle will be extremely energetic in the jet frame, and immediately attains an energy of a PeV. The associated neutrinos would come within an order of magnitude of that energy ($\sim 100$ TeV), providing a harder spectrum than the one expected from shock acceleration. 



It is expected that in the next few years neutrino telescopes will be sufficiently sensitive to test and distinguish between GRB models with different physics, and to constrain the parameters of such models. The current non-detection of neutrinos by IceCube (\citet{Abbasi:2012zw}; see also Sec. \ref{subsec:HENTel})  already questions the viability of models in which ultra high energy cosmic rays are the decay products of neutrons that have escaped the fireball with high energy. The current upper limit is still consistent with the ``standard" (i.e., following \citet{WBfireball}) predictions of neutrino emission from GRB fireballs \cite{Li:2011ah, Hummer:2011ms}. Given current uncertainties, significant constraints on this model will be obtained within 5-10 yrs of full IceCube operation, as can be seen from Fig.~\ref{fig:grb} (right panel).

\bigskip
While gamma-ray and HEN emissions from GRBs are related to the mechanisms driving the relativistic outflow, GW emission is closely connected to the central engine and hence to the progenitor of the GRB. Short-hard GRBs are thought to be driven by neutron star--neutron star or neutron star--black hole mergers\footnote{This mechanism has been described e.g. by~\citet{Eichler:1989ve}; \citet{Kochanek:1993mw}; \citet{Nakar06}; \citet{Bloom07}; \citet{Lee07}; \citet{Etienne2008}.}. GW detectors can ideally observe those binary systems up to a distance of $\sim$ 30 Mpc and $\sim 440$ Mpc for initial and advanced detectors respectively \cite{rates}. These distances coincide with the range where the HEN flux is thought to be large enough for detection with current HEN detectors. 
Note that these short GRBs are beamed and so is the expected HEN emission.  Hence one can expect cases in which GWs will be observed from such sources without an observed GRB or HEN signal. However, an orphan afterglow~\cite{Levinson:2002aw,Nakar:2002un}, macronovae~\cite{Li:1998bw} or  radio flares~\cite{Nakar:2011cw} might be observed in these cases.

Weaker GW signals are expected in any source that accelerates relativistic jets~\cite{Piran:2002kw}. 
A few mechanisms have been suggested for GW generation in long-soft GRBs, which arise during the collapse of a massive star; see e.g. \citet{Woosley2006} and references therein. 
According to the  {\it collapsar} model proposed by~\citet{Woosley99} a long GRBs arises when a relativistic jet, produced by an central inner engine penetrates the stellar envelope  of the collapsing star. The inner engine driving the jet can be either an  accreting newborn black hole~\cite{Woosley99} or  a newborn rapidly rotating magnetar \cite{Usov:1992zd}.
For accretion models, the high
rotation rate required to form the accretion disk that powers the GRB  may also lead to the 
 production of GWs via bar or fragmentation instabilities in the accretion disks and 
 also via the precession of the disks due to general relativistic effects\footnote{See e.g.~\citet{1998ApJ...502L...9F}; 
\citet{2002ApJ...579L..63D,2002ApJ...565..430F}; \citet{2003ApJ...589..861K}; 
\citet{2007ApJ...658.1173P}, \citet{Romero2010A&A}, \citet{Sun:2012bf}.}. 
Asymmetrically infalling matter produces the  burst GW signals not only at the moment of the core bounce when the central 
density exceeds nuclear density~\cite{Kotake06,Ott2008}, but also at 
the moment of the black hole formation, followed by the subsequent ring-down phases 
\cite{Ott:2010gv}. Optimistic estimates based on semi-analytical calculations suggest that the GW signals from some of these mechanisms 
are high enough to be visible in Advanced LIGO class detectors 
up to a 100 Mpc distance scale; see collective
 references in~\citet{Kotake:2012it}. However, to obtain more quantitative predictions, 
full 3D simulations using general-relativistic magnetohydrodynamics and 
sophisticated neutrino transport schemes are needed. Much effort has been recently dedicated to this issue with encouraging results; see. e.g., \citet{Mueller:2012is,Ott:2012kr,Kuroda:2012nc}, and \citet{Kotake:2012nd} for a review. Unfortunately, such studies suggest a very weak GW emission and, given the fact that the long GRB population is distributed over cosmological distances, seriously challenge the prospects for long GRB detection even in the next generation of GW detectors.
 

``Low-luminosity GRBs" ({\it ll}GRBs) are a subclass of long-soft GRBs.  {\it ll}GRBs are characterized by luminosities lower by a few orders of magnitude than typical  gamma-ray luminosities,  a smooth, single-peaked light curve, and a soft spectrum. These bursts are associated with particularly energetic  type Ibc core-collapse supernovae as observed in GRB~980425/SN~1998bw~\cite{Galama:1998ea,1998Natur.395..663K}, GRB~031203/SN~2003lw~\cite{2004ApJ...609L...5M,2004Natur.430..648S}, and GRB~060218/SN 2006aj~\cite{Campana:2006qe,2006ApJ...645L.113C,2006Natur.442.1011P,2006Natur.442.1014S}. It is interesting to note that most of the GRB-SNe associations are with {\it ll}GRBs and not with regular long GRBs.
Less luminous than typical long GRBs, these events are (not surprisingly) discovered at much smaller distances (SN~1998bw at redshift $z = 0.0085$, about 40~Mpc away from Earth; SN~2003lw at $z = 0.105$, and SN~2006aj at $z = 0.033$).
Remarkably, the event rate of {\it ll}GRBs {per unit}  local volume is more than one  order of magnitude larger than that of conventional long GRBs; see e.g.~\citet{2005MNRAS.360L..77C}, \citet{Guetta:2006gq}, \citet{2006Natur.442.1014S}, \citet{Daigne:2007qz} and \citet{Liang2007}. This makes this source population an interesting target of study from the GW+HEN point of view as well, as discussed by~\citet{Razzaque:2004yv}, \citet{2006ApJ...651L...5M}, \citet{2007APh....27..386G}, and \citet{2007PhRvD..76h3009W}.  

\citet{Bromberg+11} have recently argued that, given their apparently low power, these {\it ll}GRBs cannot arise from the regular collapsar model because the time needed for the jet to bore an escape channel through the host envelope would, in most reported cases, exceed the GRB duration. Rather, they may be gamma rays from shock break-out imparted to the  host envelope by jets that fail to emerge (``choked jets")\footnote{Relevant references include~\citet{MacFadyen:1999mk}; \citet{2001ApJ...551..946T}; \citet{Campana:2006qe}; \citet{Wang:2006jc}; \citet{Waxman:2007rr}; \citet{Katz:2009jd}; \citet{Nakar:2011mq}.}. The smooth light curve and soft spectra of these events are indeed expected from shock breakout; see~\citet{Waxman:2007rr,Katz:2009jd,Nakar:2011mq}.  Other suggested models, which produce smooth, soft emission, include scattering of the gamma-rays off an accelerating envelope or wind material \cite{1999ApJ...521L.117E}, or  gamma rays that are released from baryon-rich jet material (dirty fireballs) only after some adiabatic loss~\cite{Mandal:2010mr}. It has also been suggested that {\it ll}GRBs  are associated with the formation of magnetars rather than black holes, as argued for GRB060218 by~\citet{2006Natur.442.1018M}, a scenario that might give rise to somewhat longer GW signals~\cite{2009ApJ...702.1171C,Piro:2011ed}.  



Regardless of the question of whether or not they produce  {\it ll}GRBs, such ``choked jet" events   are interesting objects on their own, as pointed out e.g. by~\citet{1999ApJ...521L.117E}, \citet{2001PhRvL..87q1102M} and \citet{2005PhRvL..95f1103A}. In fact, late-time radio emission of some type Ic supernovae indeed suggests the presence of mildly relativistic outflow~\cite{soder,granot,mazzali,2010Natur.463..513S} that may indicate the activity of a jet in these cases.  
The expected overall energy budget of choked jets is comparable to the energy budget observed in regular GRBs\footnote{Note that if indeed a choked jet produces a  {\it ll}GRB via shock breakout, the prompt $\gamma$-rays involve only a small fraction of the total energy~\cite{Bromberg+11,Nakar:2011mq}.}. Therefore, these choked jets  could produce  GWs and HENs\footnote{The production of HEN is closely related to the efficiency of proton acceleration inside the jet, an issue which is still debated considering that the relevant shocks in these choked GRBs are expected to be radiation-dominated~\cite{Levinson:2007rj}. } in a way comparable with regular GRBs; see~\citet{1999ApJ...521L.117E}, \citet{2001PhRvL..87q1102M}, \citet{2005PhRvL..95f1103A}, \citet{Koers:2007je}, \citet{Horiuchi:2007xi}. As they are likely to be more numerous than regular GRBs (as is for example the case of {\it ll}GRBs),  they could be observed from nearer distances. In such a case, optimistic estimates predict potentially detectable levels of both GW and HEN signals, and an observable occurrence rate in the volume probed by current GW and HEN detectors. 
For example, according to~\citet{2005PhRvL..95f1103A}, an ejected mass with a kinetic energy of $3\times 10^{51}$ erg and a Lorentz factor of 3 at 10 Mpc would  generate $\sim$30 neutrino events detected in a km$^3$ detector.  These events should  be seen accompanying some specific local core collapse SNe. In this context, HEN and GW could play an interesting role in revealing the inner ``choked jets" nature of these sources.


\subsection{Bounds on the GW+HEN time delay}
  \label{sec:timewindow}

\begin{figure}
\begin{center}
\resizebox{0.8\textwidth}{!}{\includegraphics{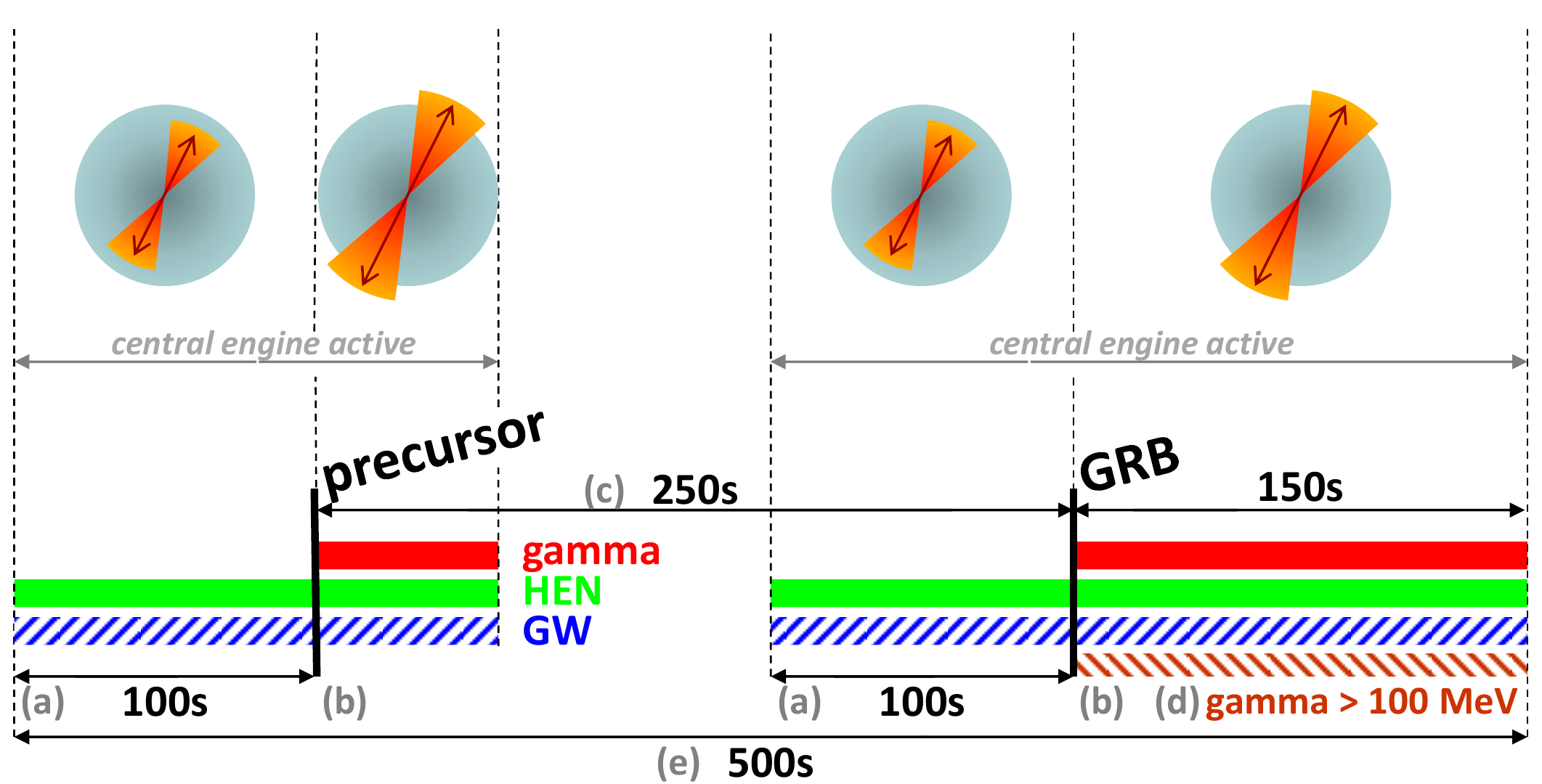}}
\end{center}
\caption{Summary of the upper bounds on the duration of GRB emission processes  taken into account in the total GW+HEN coincidence time window. (a) active central engine before the relativistic jet has broken out of the stellar envelope (note that recent estimates of jet propagation within a stellar envelope suggest that this phase lasts typically 10-20 sec; see~\citet{Bromberg:2011fg,Bromberg:2011wb}); (b) active central engine with the relativistic jet broken out of the envelope; (c) delay between the onset of the precursor and the main burst; (d) duration corresponding to $90\%$ of GeV photon emission; (e) time span of central engine activity. 
The top of the figure shows a schematic drawing of a plausible emission scenario. Figure taken from \citet{Baret20111}. }
\label{fig:emission}
\end{figure}

The possible time delay between the arrival of GWs and HENs from a given source defines the coincidence time window to apply in a multimessenger search algorithm. This window should not be too small, which could lead to the exclusion of potential emission mechanisms, nor too large, which would decrease the detection sensitivity by including non-physical coincidences.
Upon detection, the difference between the times of arrival of GW and HEN signals can give us important clues about the emission mechanism. For instance detecting a HEN prior to a GW signal may indicate that the strongest GW emission from the source is not connected to the onset of the activity of the central engine that one might expect from core-collapse models.

\citet{Baret20111} used model-motivated comparisons with GRB observations to derive a conservative coincidence time window for joint GW+HEN searches. Various GRB emission processes were considered, assuming that GW and HEN emission are connected to the activity of the central engine. Considered processes include prompt gamma-ray emission of GRBs, with a duration upper limit ($\sim 150$~s) based on BATSE observations \cite{BATSE4B}, as well as GRB precursor activity, with an upper limit on the time difference (as compared to the onset of the main burst) of $\sim 250$~s, following the analysis of \citet{PrecursorBATSE}. Further processes considered include precursor neutrino emission, as well as $\gtrsim 100$~MeV photon emission from some GRBs, as detected by Fermi LAT \cite{Atwood:2009ez}. The authors conclude that GW and HEN signals are likely to arrive within a time window of $\pm\ 500$~s, as illustrated in Fig. \ref{fig:emission}.

The time-delay between HENs and GWs could be much smaller for binary mergers which are often mentioned as the possible progenitor of short-hard GRBs. 
The amount of accreted/ejected matter involved in such case is very small, and the outflowing matter can expand unhindered, adding almost nothing to the time delay.
A semi-analytical description of the final stage of such mergers indicates that most of the matter is accreted within 1 second \cite{davies:2005}, and numerical simulations of the mass transfer suggest time scales of milliseconds \cite{shibata:2007} to a few seconds maximum \cite{Faber:2006tx}.
Therefore, the GW signal is expected to arrive very close to HENs. A window of $[-5, +1]$ seconds around the trigger time, as used for (short) GRB-GW searches in~\citet{GRB070201} and~\citet{LIGO:2010uf}, seems reasonable. 

\section{GW and HEN detection: status and prospects}
\label{sec:detection}

\subsection{Interferometric Gravitational Wave detectors}

\subsubsection{Detection principle and state of the art}

The first generation of interferometric GW detectors included a total of six
large-scale instruments. The US-based Laser Interferometer Gravitational-Wave
Observatory (LIGO, see~\citet{Abbott:2007kv}) was comprised of three kilometer-scale instruments located in Livingston, Louisiana and Hanford, Washington (the latter hosted two interferometers in the same vacuum enclosure). The French-Italian project Virgo~\cite{accadia12:_virgo} had one instrument of the same class located in Cascina near Pisa, Italy. This
set of kilometer-scale instruments was complemented by a couple of detectors with
more modest dimensions (several hundreds of meters): GEO~\cite{Grote:2010zz}, a German-British
detector in operation near Hanover, Germany and the Japanese prototype CLIO~\cite{Agatsuma:2009tm} located in the Kamioka mine.


\begin{figure}
\includegraphics[width=0.6\linewidth]{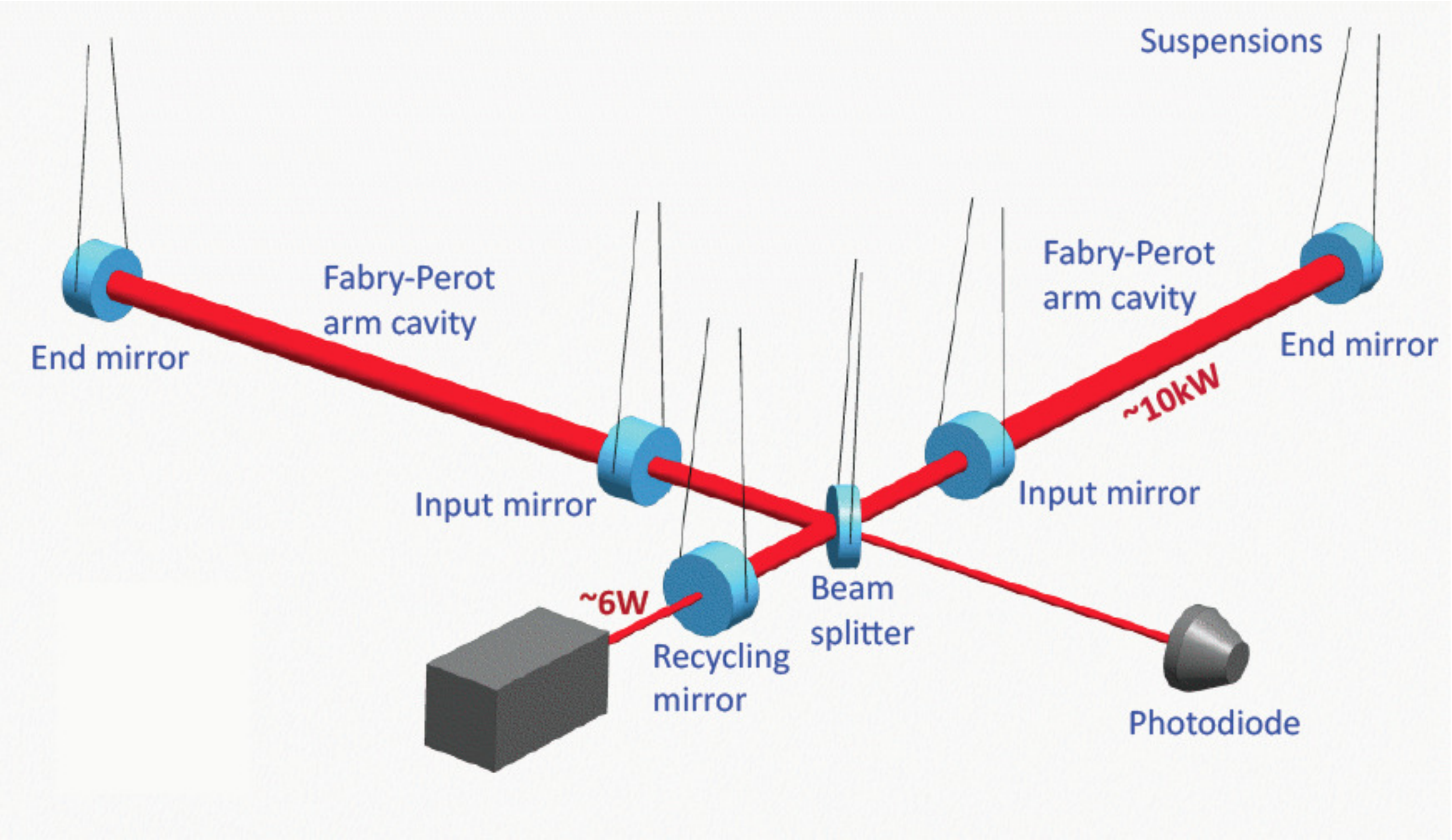}
\caption{Optical scheme of a gravitational-wave interferometric detector, consisting of two twin laser beams propagating in km-long arms oriented at 90$^\circ$ to each other. The Fabry-P\'erot cavities enable the storage of the beams, thereby increasing by a significant factor their effective path length; the suspended, highly reflective mirrors play the role of test masses.  A gravitational wave would cause a difference in the optical pathlengths in each arm, which can be inferred by measuring the interference pattern at the photodiode. Figure adapted from Y. Aso (GECo, Columbia University).}
\label{fig:GWInterferometer}
\end{figure}

\begin{figure}[!t]
\includegraphics[width=0.45\linewidth]{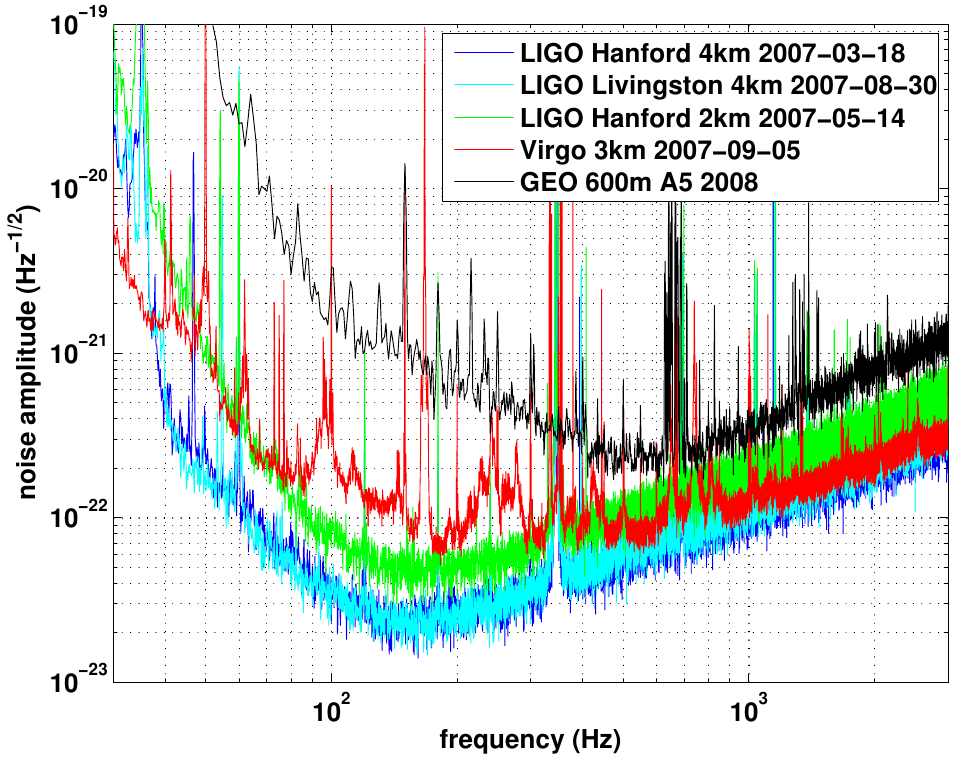}
\hfill
\includegraphics[width=0.52\linewidth]{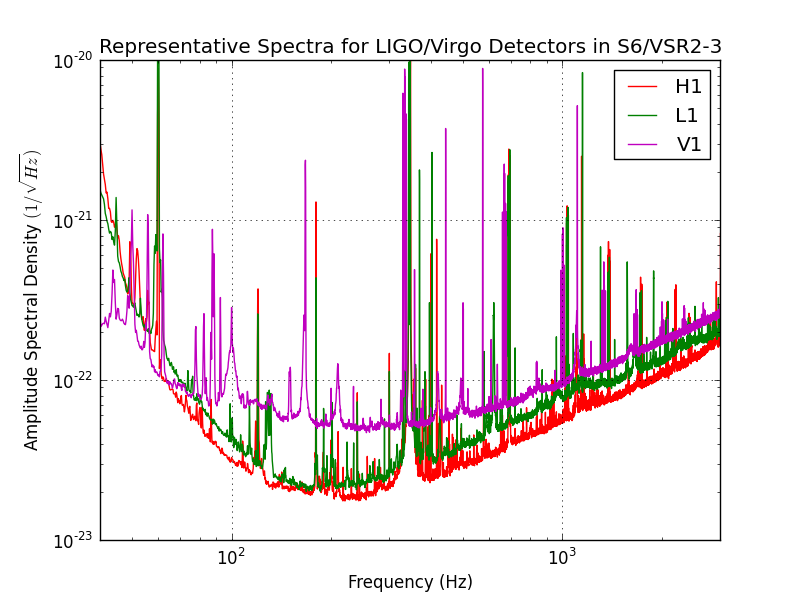}
\caption{Detector noise spectra from LIGO and Virgo associated with the two joint science runs, showing the increase of sensitivity of the detectors. Left panel shows the best sensitivity  associated with S5/VSR1 data set (2007), along with that of GEO (2008), as taken from~\citet{Abadiemagnetar}; right panel shows typical sensitivities for LIGO and Virgo associated with S6/VSR2-3 (2009--2010) data sets, as from~\citet{LIGO:2012aa}.} 
\label{fig:gwnoise}
\end{figure}

Despite major differences in the technologies, all past and upcoming ground-based km-scale detectors measure gravitational waves through the same principle (see
Fig. \ref{fig:GWInterferometer} for an illustration of the general scheme). They
all sense the strain that a passing GW exerts on space-time by monitoring the
differential length $\delta \ell$ of the optical path followed by two laser
beams propagating along orthogonal directions. Measurement noises (mainly the
thermal noise due to the Brownian agitation of the atoms constitutive of the
main optics and the shot noise due to the quantum nature of light) can be
reduced to reach the level of $h \equiv \delta \ell/L \sim 10^{-21}$, where $h$
is the GW amplitude and $L$ is the total optical path length. This best
sensitivity was achieved in a frequency band ranging from approximately $\sim 100$ Hz to $1$
kHz (see Fig.~\ref{fig:gwnoise}) and approaches the optimistic 
theoretical expectations from the astrophysical sources presented earlier.

\begin{figure}
\begin{center}
\resizebox{0.7\textwidth}{!}{\includegraphics{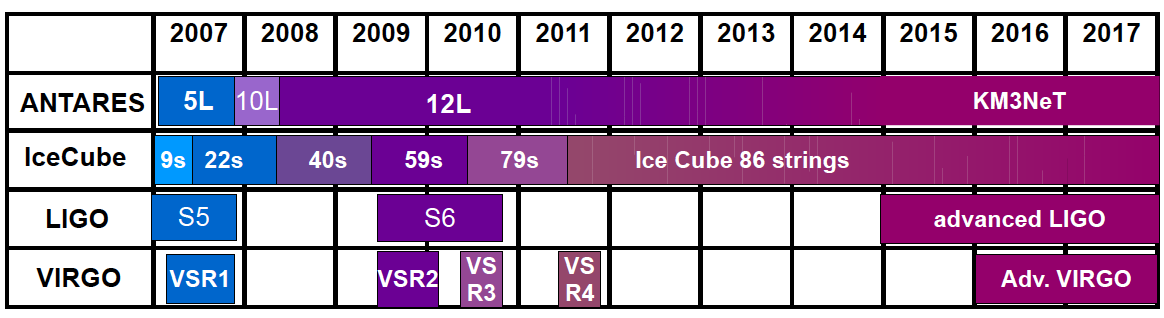}}
\end{center}
\caption{Time chart of the data-taking periods for the ANTARES (and KM3NeT), IceCube, LIGO and Virgo experiments, indicating the respective (achieved or planned) upgrades of the detectors. The IceCube detector is now complete and will be operating for at least another 5 years, with possible upgrades in the meantime. The deployment of the KM3NeT neutrino telescope, which will take place in parallel with the operation of ANTARES, is expected to last three to four years, possibly starting in 2014. The detector will be taking data with an increasing number of photodetectors before reaching its final configuration. A larger-scale upgrade to the next generation of GW interferometers (advanced LIGO and Virgo) is ongoing and data taking should start again around 2015.} 
\label{fig:chart}
\end{figure}

The initial LIGO and Virgo detectors have conducted several campaigns of joint data taking (``science runs"), as illustrated in Fig.~\ref{fig:chart}. These data have been searched for a broad range of GW signatures. Those signatures are either from short transient
sources associated with very energetic cataclysmic events like mergers of
neutron star and/or black hole binaries, or from long-lived permanent
sources such as deformed neutron stars or stochastic backgrounds resulting from
the superposition of many unresolved sources. No gravitational wave has been
detected so far, however, interesting upper limits were placed on the GW strain amplitude
from a variety of targeted sources. We will focus here on the first category
(transients) since it pertains to the main interest of this paper.

The two joint science runs conducted by LIGO and Virgo were labelled S5 for LIGO and VSR1 
for Virgo for the first run, and S6 and VSR2/3 for the second. A total of 635  
days of observing time have been analyzed~\cite{reference1,Abadie:2012rq}. The upper limit (at 50\% confidence level) on the GW strain obtained from an all-sky all-time search
is slightly below $h_{rss} \lesssim 5 \times 10^{-22}$ Hz$^{-1/2}$ for waveform
frequency at about 200 Hz, where the bound is on the root-square-sum amplitude
$h^2_{rss} \equiv \int dt\: h^2_+(t) + h^2_\times(t)$ of the two GW
polarizations, $h_+$ and $h_\times$, at Earth. Note that the exact result
depends on the assumed GW model (the generic choice considered here are sine
Gaussian waveforms of various central frequencies).  Assuming a linearly
polarized wave and averaging over the inclination of the source, this strain
limit corresponds to a GW burst energy of $2 \times 10^{-8} M_{\odot} c^2$ for a source
at Galactic distance of $10$ kpc, and $5 \times 10^{-2} M_{\odot} c^2$ for a
source located in the Virgo cluster (at a distance of 15 Mpc). Those estimates
are comparable to the expected GW-radiated energy from core-collapses and
mergers of stellar-mass compact objects respectively. The same data, when
searched specifically for inspiraling binaries of neutron stars, led to an upper
limit on the rate of such astrophysical events of ${\cal R}_{90\%}=1.3 \times
10^{-4} \mathrm{yr}^{-1} L_{10}^{-1}$ \cite{reference2,PhysRevD.85.082002} which is still two orders of magnitude larger than the rate estimate obtained from population models
\cite{rates}.


The building of next generation of GW instruments is underway, with data collection expected to start  around 2015. Following~\citet{rates}, source population models imply that direct detection of gravitational waves can be achieved within the next decade by such advanced ground-based GW detectors: advanced LIGO\footnote{\url{https://www.advancedligo.mit.edu}} in the USA~\cite{Harry:2010zz}, Advanced VIRGO\footnote{\url{http://wwwcascina.virgo.infn.it/advirgo/}} in Italy and KAGRA in Japan~\cite{0264-9381-27-8-084004}. There is now a proposal (still to be approved by U.S. and Indian institutions) to move one of the three advanced LIGO detectors at a new observatory in India. If this plan materializes, the detector at the Indian site would start operation around 2020.  
These detectors will offer a tenfold sensitivity increase over the initial detectors  around 100Hz and will extend the observable frequency range by almost a decade down to 10Hz.
Observation-based models predict e.g. a detectable rate for binary neutron star coalescence between about 0.4 to 400 events annually~\cite{rates}.

\subsubsection{Multimessenger strategies}
\label{MMGW}


The GW+HEN program discussed here is one example of a number of efforts to develop joint analyses targeting GWs and other cosmic messengers. Another class of examples are the searches for GW transients in coincidence with (or
``triggered'' by) the high-energy photons from (short and long) GRBs (see
e.g. \citet{Abbott:2009kk}, \citet{LIGO:2010uf} and references therein) and SGRs
(see e.g. \citet{Abadiemagnetar} and references therein). Since, in this case, the event time and
the sky direction of the source are known, a more sensitive and narrowly focused search can be conducted relative to the aforementioned all-sky searches. Two cases, GRB 070201 and
GRB 051103, have received particular interest \cite{GRB070201,Abadie:2012bz}.
The error box on the location of those two GRBs overlaps with close galaxies M31 and M81
respectively. The analysis of GW data in coincidence with both GRBs yielded a
null result. Under the assumption that the progenitors of those events are
located in the identified host galaxies, binary mergers (or any other source
emitting more than $\sim 10^{-4} M_\odot c^2$ as GW transients) are excluded with
high confidence from this non-detection statement. It is also worth mentioning
that an electromagnetic follow-up program of candidate GW triggers has been
performed during S6--VSR2/3 \cite{lsc:_implem}. This program involved a range
of robotic telescopes including the Liverpool Telescope, the Palomar Transient
Factory, Pi of the Sky, QUEST, ROTSE, SkyMapper, TAROT and the Zadko Telescope
observing the sky in the optical band, the {\it Swift} satellite with X-ray and
UV/Optical telescopes and the radio interferometer {\it LOFAR}.


For more details on direct detection of gravitational waves and its implications
for astrophysics and cosmology, we refer the reader to
\citet{sathyaprakash09:_physic_astrop_cosmol_gravit_waves}.

\subsection{High-energy neutrino telescopes}
\label{subsec:HENTel}

\subsubsection{Detection principle and state of the art}

\begin{figure}[!t]
\includegraphics[width=0.49\linewidth]{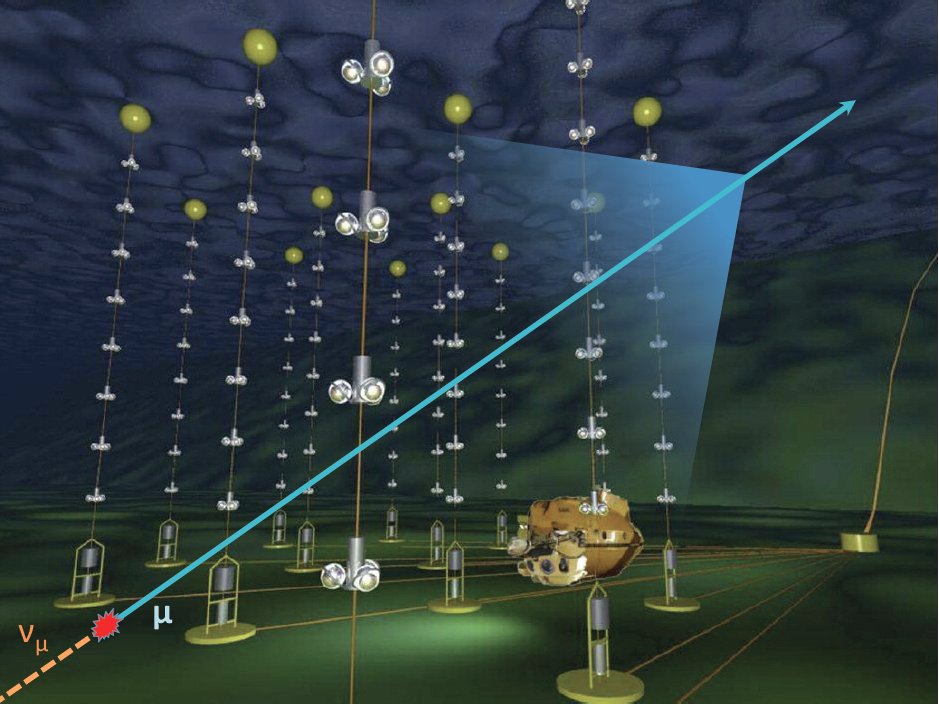}
\hfill
\includegraphics[width=0.50\linewidth, height=7cm]{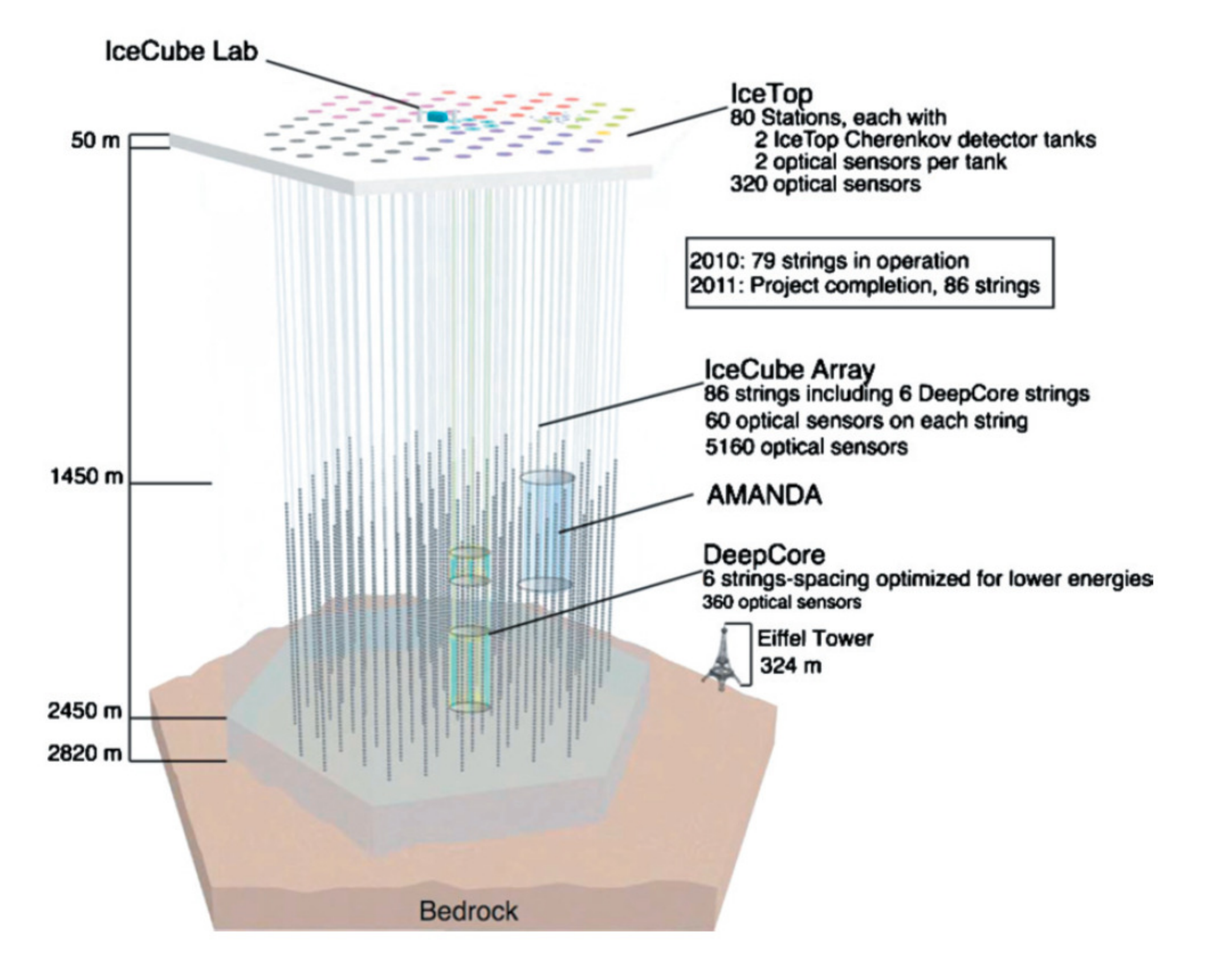}
\caption{\textbf{Left:} Schematic view of the ANTARES detector, with an illustration of its detection principle: an upgoing $\nu_\mu$ interacting with matter in the seabed will produce a muon that can cross the detector, emitting a cone of Cherenkov light that will be detected by the array of photosensors.  ANTARES consists of 12 instrumented lines anchored to the seabed at a depth of 2475 m, about 40 km off the coast of Toulon (france). The lines are 450 m long and kept up by a buoy; each of them supports 25 detection storeys separated by 14.5 m and supporting triplets of optical modules looking 45$^\circ$ downwards.  The inter-line separation is about 70 m. Adapted from http://antares.in2p3.fr.\textbf{Right:} Schematic view of the IceCube detector at the South Pole. The strings are deployed from the surface and instrumented between 1450 m and 2450 m; every  string supports 60 optical modules enclosing one downward-looking PMT each. The inter-line separation is about 125 m. The sketch also shows the position of the prototype detector AMANDA, of the low-energy infill DeepCore, and of the surface array IceTop. From http://icecube.wisc.edu.}   
\label{fig:nutel}
\end{figure}

Given the very weak neutrino cross section and the 
typical astrophysical spectra falling as a power-law at high
energies, HEN astronomy requires the instrumentation of huge ($\sim 1$ km$^3$) volumes
of target material.
The concept of neutrino telescopes appeared in 1961 when M.A. Markov
proposed to use the water of deep lakes or the sea to detect the secondary
muons created in the charged-current interaction of HEN with nuclei. The Cherenkov light emitted by the muon in a transparent medium can be used to
infer the arrival direction of the neutrino \cite{markov}. This detection principle takes advantage of the fact that the muon track
can be several kilometers long, thus enhancing the effective volume of
the detector. Such neutrino telescopes have been built in the form of three-dimensional arrays of photomultiplier tubes (PMTs) embedded in pressure-proof glass spheres arranged on vertical cable strings, with an inter-storey spacing of a few tens of meters and an inter-string distance up to about 100 meters; see Figure~\ref{fig:nutel} for an illustration of this detection principle.
The knowledge of the timing and amplitude of the light pulses recorded by the PMTs allows  reconstruction of the trajectory of the muon, as well as inference of the arrival direction of the incident neutrino.

These detectors have to cope with a large background 
of high-energy muons originating in the air showers generated by the interaction of high-energy cosmic rays with the atmosphere. They are therefore  installed beneath
thousands of meters of water-equivalent shielding, restricting the possible sites to 
deep lakes, the deep sea, or the south pole glacier.
Even with this shielding, the rate of atmospheric muons is several
orders of magnitude above the rate of genuine atmospheric neutrinos, i.e. neutrinos created in cosmic-ray interactions in the atmosphere. To further reduce this background, the detectors are optimized to detect up-going muons produced by neutrinos that have traversed the Earth (which acts as a shield against all other particles).  The field of view of neutrino telescopes is therefore  $2\pi$ sr for neutrino energies
100 GeV  $\leq E_\nu \leq 100$ TeV; a detector placed in
the southern (resp. northern) hemisphere will observe the northern (resp. southern) sky. Above this energy, the sky coverage is reduced because of neutrino absorption in the Earth. However, it can be partially recovered by looking for horizontal and downward-going neutrinos, which can be more easily separated from the background of atmospheric muons because of their much higher energies.

 
 Three neutrino telescopes are currently operating worldwide. The most advanced is IceCube~\cite{icecubeGen}, which has recently achieved its final configuration with 86 strings, instrumenting one km$^3$ of South pole ice at depths between 1500 m and 2500 m (see right panel of Fig.~\ref{fig:nutel}). IceCube possesses a denser infill of 6 additional strings dubbed as DeepCore~\cite{Collaboration:2011ym}, which extends its detection capabilities at lower energies (down to $\sim$ 10 GeV). It is also complemented on the surface by a 1 km$^2$ air shower array named IceTop~\cite{Stanev:2009ce}, which is used for CR composition studies, and in coincidence with the in-ice detector. 
Another neutrino telescope has been operating for some years in Lake Baikal~\cite{baikal}
in a much smaller configuration; it has recently deployed 3 prototype strings for a km$^3$ scale detector.
Finally, ANTARES~\cite{antaresGen} is a neutrino telescope deployed at depths from 2000 m to 2500 m in the Mediterranean Sea,  near Toulon
(France); it has been operating in its complete, 12-line configuration since mid-2008 (see left panel of Fig. \ref{fig:nutel}). ANTARES has been joined by the two prototype projects NEMO \cite{Taiuti:2011zz} and NESTOR \cite{nestor} in forming the European Consortium KM3NeT which aims at the construction of a km$^3$ scale telescope in the Mediterranean, whose operation could start in 2014~\cite{Katz:2011zz}. This second kilometer-scale project would allow all-sky coverage, and in particular the monitoring of a large fraction of the Galactic Plane, which contains many potential sources. An interesting characteristic of these detectors is the ability to take data during the construction phase; each data taking configuration is labelled by the number of functioning detector lines (or strings). 
 Detector
performance, as measured by e.g. effective area and pointing
accuracy, then improves as new lines are added. The time chart in Fig.~\ref{fig:chart} presents the different phases of operation of the IceCube and ANTARES telescopes from 2007 onwards. 

\begin{figure}
\includegraphics[width=0.6\linewidth]{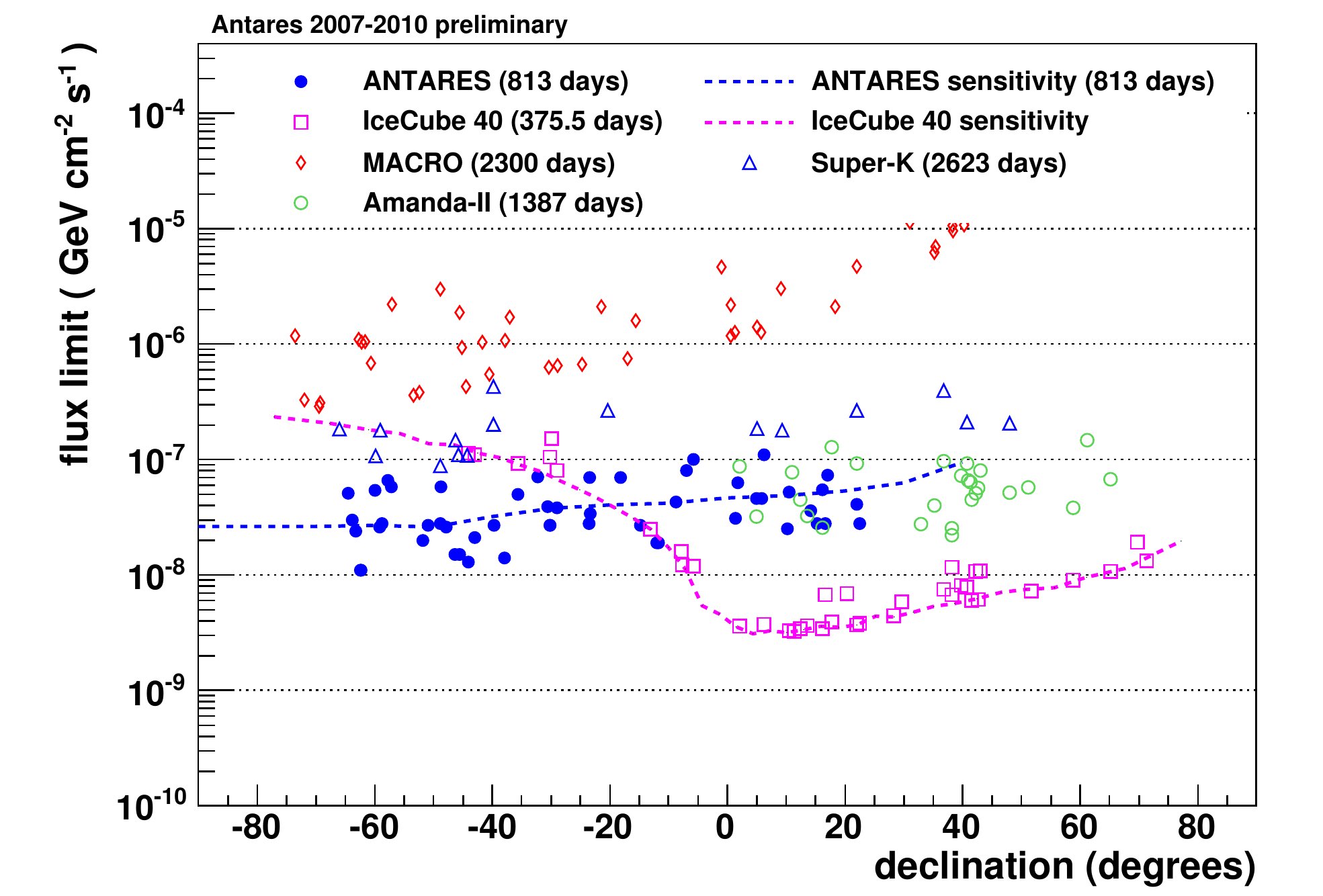}
\caption{Current experimental sensitivities to differential cosmic neutrino flux $E^{-2}\frac{dN}{dE}$ (solid lines) and limits (points) as a  function of declination for time-integrated searches of point-like sources. The figure shows the results of AMANDA-II~\cite{amandaPS} and its successor IceCube, in its  
40-string configuration~\cite{PS_IC40}, as well as those of ANTARES; also displayed are the limits from the MACRO~\cite{macro} and Super-Kamiokande~\cite{sk} experiments, which are not principally devoted to neutrino astronomy. Figure taken from~\citet{AdrianMartinez:2011ax}. }
\label{ps_results}
\end{figure}

No evidence for a cosmic neutrino has been reported so far by any of these detectors. Limits have been set both for point sources~\cite{PS_IC40,AdrianMartinez:2011ax} and for the diffuse all-sky flux expected from the large-scale distribution of sources which are individually too faint to resolve~\cite{:2010ec,Abbasi:2011jx}. 
Figure \ref{ps_results} presents the latest results of ANTARES and IceCube searches for point-like sources of high energy neutrinos, showing the significant increase in sensitivity with respect to the previous generation of detectors such as SuperKamiokande and MACRO.

\subsubsection{Multimessenger strategies}
\label{MMHEN}

A wide variety of generic and specialized searches is also performed on the neutrino data, many of which make use of time-dependent observations from photon experiments.  Searches have been performed e.g. for neutrinos in coincidence with  flares from active galactic nuclei~\cite{:2011ara,AdrianMartinez:2011jz} and from the Crab nebula \cite{IceCube:2011aa}, for periodic neutrino emission from binary systems \cite{Abbasi:2011ke} and for neutrinos from GRBs \cite{Abbasi:2009kq,Abbasi:2011qc}.  

The IceCube collaboration has recently published the most stringent limit to date on HEN emission in coincidence with GRBs \cite{Abbasi:2012zw}. Two differently optimized analyses have been performed using two years of data collected by the partially-constructed detector between 2008 and 2010. No candidate neutrino events were found in either analysis.  This yields upper limits below many earlier model predictions, in particular those postulating that GRBs are the source of UHECRs, making it possible to put strong constraints on model parameters like the jet Lorentz factor or the ratio of energy in protons to the energy in leptons in the jet.
As stated in Sec.~\ref{sec:grb}, it is important to note that this does not rule out the fireball phenomenologically; rather, it shows that detectors are now becoming sensitive enough to test many existing models and constrain their parameters. Furthermore, neutrinos that are predicted in afterglow models at higher energies and in precursor models at lower energies are not yet constrained by existing data. 

Additionally, the near-simultaneous arrival of two or more neutrinos from the same direction could indicate that a highly energetic burst has occurred.  If this is detected in real-time, then neutrino telescopes can be used as triggers for optical, x-ray, and gamma-ray follow-ups.  IceCube  and ANTARES  currently have alert programs established or in development with fast optical telescope networks like ROTSE and TAROT, gamma-ray telescopes such as Swift, Fermi, MAGIC, and VERITAS; see e.g.~\citet{followup}, \citet{VanElewyck:2011zz}, \citet{tatoo}. 
The ROTSE optical follow-up program implemented for neutrino multiplets detected with IceCube has e.g. been used to search for soft relativistic jets in core-collapse supernovae, yielding negative results so far \cite{Abbasi:2011ja}.

``All-sky'' instruments like neutrino telescopes and GW detectors thus provide an opportunity to alert pointing instruments before or concurrent with an interesting astronomical event, in addition to enabling more powerful joint searches for bursts in a completely offline way after the data has been recorded. It should be noted that an analogous possibility exists for much lower energy neutrinos. Because the PMT dark noise rate is particularly low in ice, the IceCube detector is sensitive to sudden fluxes of MeV neutrinos which lead to collective rise in the PMT rates.  Nearby supernova up to 50 kpc can be detected this way. IceCube is therefore part of the SuperNova Early Warning System (SNEWS), sending real-time triggers when the collective PMT rate passes a given threshold; see~\citet{Kowarik:2009qr} and \citet{IceCubeSN}.  As with high energy neutrino data, an offline analysis with other all-sky instruments, like GW detectors, is also possible. While the MeV neutrino signal does not provide any directional information, the time-coincidence search would allow exploring the data below the higher threshold required for SNEWS \cite{Halzen:2009sm}.
 
 More information on the experimental aspects and physics reach of high-energy neutrino astronomy can be found in recent reviews by \citet{chiarusi}, \citet{teresa}, \citet{rop} and \citet{Katz:2011ke}.


   %
 
\section{Perspectives for the joint data analysis}
\label{sec:gwhen}


\subsection{GW data analysis}

GW signals are expected to be infrequent and have low signal-to-noise ratios in both the current and next generation of ground-based detectors. Searching for GW transients therefore consists of searching for weak and rare signatures in long duration time-series. The identification of these
weak signals is confounded by the presence of ``glitches'': non-Gaussian, 
non-stationary fluctuations in the background noise.  Glitches are produced by a
variety of environmental and instrumental processes, such as local seismic noise
or saturations in feedback control systems.  Since glitches occasionally occur
nearly simultaneously in separate detectors by chance, they can mimic a GW
signal~\cite{2008CQGra..25r4004B}. The presence of glitches is currently the
main limiting factor to the sensitivity of searches for transient GW. In this Section, we review the
various techniques in use to discriminate between true signals and background
noise. We then examine the source localization and angular accuracy of networks of
multiple GW detectors.

\subsubsection{Multi-detector coherent analysis and background rejection}

{\em Coherent excess power} methods\footnote{See for example
  \citet{Guersel:1989th}; \citet{FlHu:98b}; \citet{AnBrCrFl:01};
  \citet{Mohanty:2006ha}; \citet{Rakhmanov:2006qm}; \citet{Ch_etal:06};
  \citet{Summerscales:2007xq}; \citet{Klimenko:2008fu}.} used to search for
unmodeled GW bursts require that the signals received by all detectors be
consistent in time \textit{and} phase. Concretely this is realized by combining
the data streams from multiple detectors, taking into account the antenna
response and noise level of each detector so that the sum operates
constructively for a GW burst from a given sky direction. The data stream which
results from this coherent combination maximises the signal-to-noise ratio (SNR).
It is used to produce a time-frequency map of the signal energy (equivalently,
the SNR), which is then scanned for transient excursions (or {\it events}) that
may be GW signals.  Each event is characterised by a measure of significance,
based on energy and/or correlation between detectors, as well as its
time-frequency properties.

It is also possible that cross-correlating data from multiple detectors can result in interference between the GW signals, suppressing the signal rather than background glitches. The energy in these ``null'' stream(s) may be used to reject or down-weight events that are inconsistent with a
gravitational wave. The success of such tests depend critically on having
several independent detectors of comparable sensitivity. Cross-correlation
with auxiliary non-GW sensitive channels (which monitors the global status and
environment of the GW detector) also provides an important resource for
background rejection.

For certain sources, the expected GW signature is known. Such morphological
information may help distinguish a real signal from background noise. This
is the case for the coalescence of two compact objects, e.g. two neutron stars
or a neutron star and a black hole. ``Template waveforms'' for such sources are
obtained from post-Newtonian approximations of the binary dynamics in their inspiral phase 
\cite{Blanchet:2002av, Buonanno:2006ui}. The search for signals with this
specific morphology is realized by matched-filter algorithms which cross-correlate the data with the templates \cite{Allen:2005fk,Harry:2011qh}.

\subsubsection{Source localization}

GW detectors are non-imaging instruments with a nearly omnidirectional response.
Source localization therefore requires multiple detectors; the relative amplitudes and time delay between signals received at different locations allows for triangulation of the source. Several methods of localisation have been investigated\footnote{See e. g.  \citet{Guersel:1989th};
  \citet{Wen:2005ui}; \citet{Cavalier:2006rz}; \citet{Rakhmanov:2006qm};
  \citet{Acernese:2007zza}; \citet{Searle:2007uv}; \citet{Searle:2008ap};
  \citet{Wen:2008zzb}; \citet{Markowitz:2008zj};
  \citet{Fairhurst:2009tc}; 
  \citet{Wen:2010cr}.}. \citet{Fairhurst:2009tc} gives the following
approximation for the timing accuracy of a GW signal: $
 \sigma_{t} \sim (2\pi \sigma_{f} \rho)^{-1} \, ,
 $ where $\sigma_{f}$ is the effective bandwidth of the source and $\rho$ is the
 SNR. For nominal values $\sigma_f = 100$ Hz and $\rho=8$, timing accuracies are
 on the order of 0.1 ms.  This can be compared to the light travel time between
 detectors, 10 -- 30 ms for the LIGO-Virgo network.  For example, for a binary
 coalescence signal at the threshold of detectability, \citet{Fairhurst:2009tc}
 estimates a best-case localization of 20 deg$^{2}$ (90\% containment), and a
 typical localization of twice this. Additional constraints provided by other
 instruments with a better angular accuracy such as HEN telescopes can therefore
 significantly help improve the source localization.


 %

\subsection{HEN data analysis}


The searches for astrophysical point-sources of HEN rely principally on charged-current interactions of muon neutrinos. At the energies probed by neutrino telescopes, the outgoing muon can travel from hundreds of meters up to many kilometers, and the direction of the muon is nearly collinear with the original direction of the neutrino.  Cherenkov photons propagating from the track are detected by the array of photo-multiplier tubes, and the relative timing of the photon hits is used to reconstruct the muon direction. The angular resolution is limited by the number of hits detected and by any distortions in the photon arrival times due to scattering in the water or ice.  Higher energy muons are generally better reconstructed, since they travel farther, providing a longer lever arm for reconstruction, and since more photons are emitted in stochastic energy losses along the path.  

The track reconstruction principle is to maximize the
likelihood of time residuals of photon hits, but this is complicated by the presence of the  abundant background of atmospheric muons. At the reconstruction level, the rate of downgoing atmospheric muons that are wrongly reconstructed as upgoing tracks 
 is still several orders of magnitude larger than
the rate of genuine upgoing muons events that come from 
atmospheric neutrinos that have traversed the Earth. This background of misreconstructed tracks can be reduced to about a few percent of the bulk of upgoing atmospheric neutrinos by applying quality cuts e.g. on the likelihood of the track. The angular resolution above 10~TeV is essentially determined by the scattering length of light in the medium,
yielding a median error angle on the neutrino direction of about
$0.1^{\circ}$ in the deep sea and $0.5^{\circ}$ in the South Pole
glacier for telescopes of km$^3$-scale size. 
 
The energy of the incoming neutrino is estimated based on the 
amount of Cherenkov light detected from the muon track.
The simplest estimator is the total number of photon hits detected
from the track. Given that only a fraction of the muon track is
contained in the instrumented volume, the resolution is intrinsically
limited and is usually a lower bound, since a muon from a high energy
neutrino interaction many kilometers away will lose a large fraction of its energy before
reaching the detector.
On the other hand, the observation of a large number of photons from a track unequivocally signals a high-energy event.
Therefore energy estimation can still be used to distinguish cosmic
neutrinos from atmospheric ones, because the atmospheric spectrum is known to fall steeply with energy ($\sim E^{-3.7}$) whereas cosmic fluxes can be much harder with
a typical $E^{-2}$ spectrum extending to PeV energies.



At energies well above a TeV, up-going tracks are generally straightforward to separate from the misreconstructed muon background.  At lower energies, the smaller number of photon hits makes it increasingly difficult to perform this separation.  Steady point source searches often place less emphasis on this low-energy range, because the steepness of the atmospheric neutrino background means that even if a high purity neutrino sample is obtained, the low energy region is dominated by the pile-up of atmospheric neutrinos.  However, on very short time scales such as $\sim$\,second-long bursts, even the atmospheric neutrino background is small.  This means that more powerful event selection methods, for instance machine learning algorithms like boosted decision trees, can be used to separate lower energy neutrino events from the mis-reconstructed muon background.  The development of these techniques will play an important role in searches for objects like choked GRBs, where the neutrino energies may be at $\sim$\,TeV and below.

\subsection{HEN-triggered GW searches}
\label{subsec:trig}

\begin{figure}
\begin{center}
\resizebox{0.6\textwidth}{!}{\includegraphics{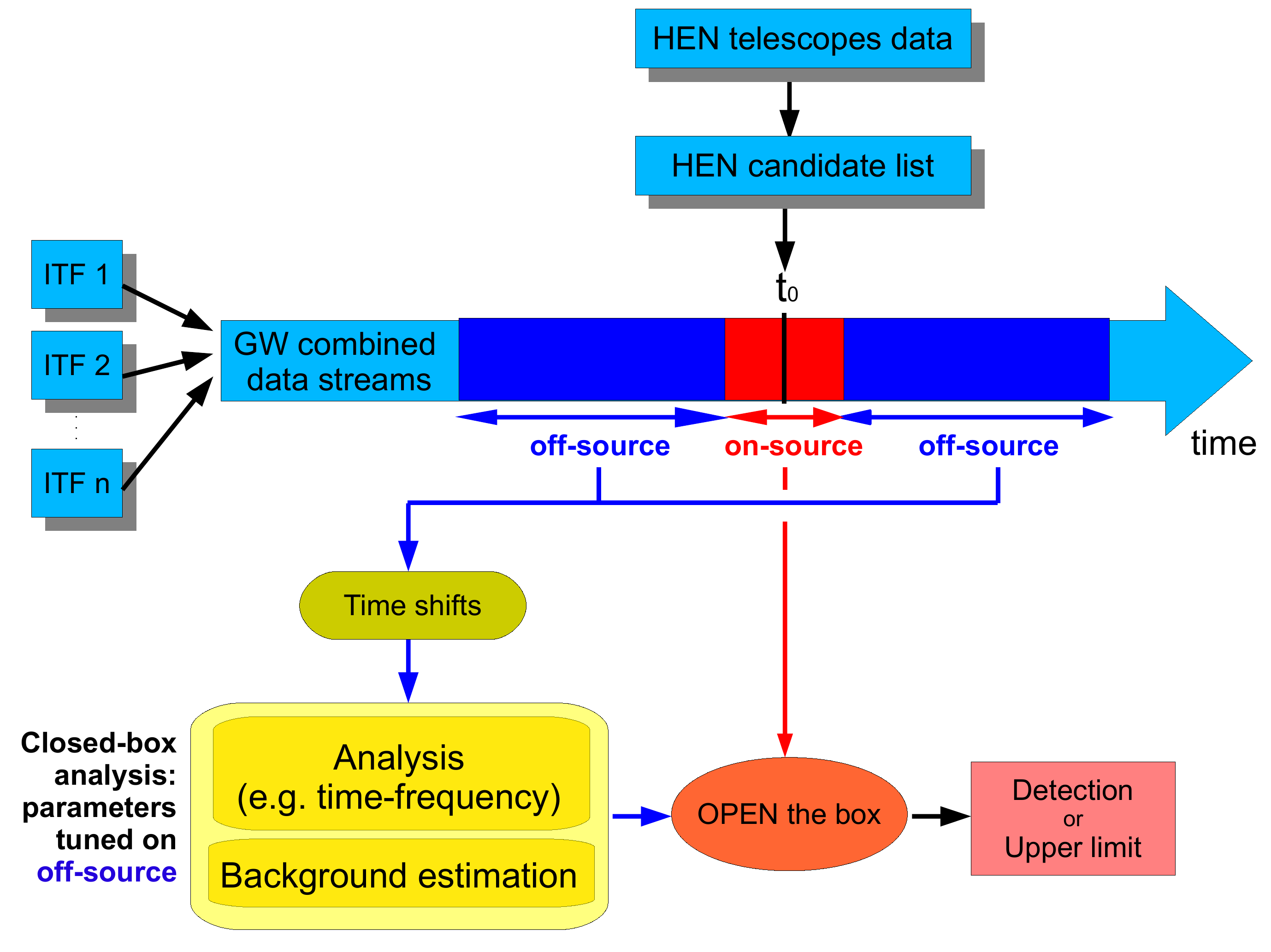}}
\end{center}
\caption{Schematic flow diagram of a HEN-triggered search for GWs. Each neutrino candidate (with its time and directional information) provided by a HEN telescope acts as an external trigger for the X-pipeline, which searches the combined GW data flow from all active interferometers (ITFs) for a possible concomitant signal. The size of the spatial search window is related to the angular accuracy associated with each HEN candidate. The background estimation and the optimization of the selection strategy are performed using time-shifted data from the off-source region in order to avoid contamination by a potential GW signal. Once the search parameters are tuned, the box is opened and the analysis is applied to the on-source data set.} \label{fig:HENtrig}
\end{figure}


One of the simplest GW+HEN coincidence searches that may be performed would search the GW data around the neutrino arrival time in the estimated direction of the neutrino candidate. 
Thanks to the reduction in the volume of analyzed data, such triggered GW searches can be run with a lower event detection
threshold than an un-triggered search, leading to a higher detection probability at a fixed false alarm probability and better limits in the absence of detection.
Similarly, the \textit{a priori} knowledge of the source direction allows for searching only a small part of the sky and veto candidate events seen in multiple detectors at times
not consistent with the expected GW arrival time difference. In fact, the number of accidental coincidences between GW detectors decreases with the
size of the search time window. Thus, the use of an external trigger can be a very effective tool for a successful search of GW signals. 
A variety of algorithms for triggered searches have been developed. We will focus here on two examples which are currently used in the context of HEN-triggered searches: the burst search algorithm dubbed as X-pipeline~\cite{Sutton:2009gi}, and the STAMP search algorithm for extended GW emission~\cite{thrane11:_long}. 

 The X-Pipeline is a coherent technique that has been used to perform searches for GWs in
association with GRBs~\cite{Abbott:2009kk}. It is a software package designed
for autonomous searches for unmodelled GW bursts. It
targets GW bursts associated with external astrophysical ``triggers" such as GRBs or neutrinos.  It performs a
coherent analysis of data from arbitrary networks of GW
 detectors, while being robust against noise-induced glitches. This allows the
analysis of each external trigger to be optimized independently, based on
background noise characteristics and detector performance at the time of the
trigger, maximizing the search sensitivity and the chances of making a
detection. The pipeline also accounts for effects of uncertainties in the
results such as those due to calibration amplitude, phase, and timing.

 Stochastic Transient Analysis Multi-detector Pipeline (STAMP) is a cross-correlation-based algorithm that looks for structures due to GWs in
 cross-power frequency-time maps~\cite{thrane11:_long}. Cross-power maps are produced by cross-correlating strain data from two spatially separated GW detectors after
 applying a filter function. By choosing a proper filter function, STAMP can
 search for a GW signal from a particular direction in the sky. Due to its
 cross-correlation approach, STAMP mitigates noise glitches due to environmental
 factors. In STAMP, the SNR of any GW signal will increase as $\sqrt{T}$ where
 $T$ is the duration of the signal. This makes STAMP suitable for GW signals
 with duration of tens of seconds to weeks while X-Pipeline is more commonly
 applied to signals with duration of second or less.

In this context, a neutrino source will be characterised by a set of inputs for the search algorithms:  its sky position, as given by the direction of the reconstructed muon track in the neutrino telescope, the associated (and possibly direction-dependent) point-spread function of the detector, the neutrino arrival time, which defines the trigger time $t_0$, and the range of possible time delays (positive and negative) $\Delta t$ between the neutrino signal and the associated GW signal, which is astrophysically motivated, as discussed in Sec.~\ref{sec:timewindow}. The latter quantity is referred to as the  \emph{on-source} window for the neutrino; this is the time interval which is searched for GW candidate signals.

A crucial part of the procedure is the estimation of the background distributions, which is performed on the data pertaining to an \emph{off-source} time window, typically covering about $1.5$ hours around the neutrino time trigger and excluding the on-source interval.  This strategy ensures that the background does not contain any signal associated with the neutrino event but has similar statistical features as the data searched in association with the neutrino. This time range is limited enough so that the detectors should be in a similar state of operation as during the neutrino on-source interval, but long enough to provide off-source segments for estimating 
the background. A schematic flowchart of this analysis strategy is presented in Fig.~\ref{fig:HENtrig}.

A GW burst search with the X-Pipeline was performed using a list of 158 ANTARES 5L neutrino candidates as external triggers. These triggers were obtained during the data taking period from February to September 2007, in coincidence with initial LIGO (S5) and Virgo (VSR1) detectors  \cite{AdrianMartinez:2012tf}
Each selected HEN was characterised by its time, arrival direction and angular uncertainty which defines the size of the spatial box to be searched for a GW signal \footnote{More precisely, the scan of the angular search window performed by the X-pipeline was weighted using a log-normal parameterisation of the HEN point-spread function.}. Consistently with~\citet{Baret20111}, the on-source window was defined as a symmetrical interval of $\pm 500$ s around the HEN time, in which data were searched for GW signals of different durations by means of a time-frequency analysis. The event-per-event significance of the resulting GW candidates was estimated by performing the same analysis on the off-source window, where no signal was expected. To account for the trial factor due to the number of HEN events analyzed, a binomial test was then applied to search for a cumulative excess above the expected background level, following a procedure already applied to the GRB-triggered GW searches, e.g.  in~\cite{Abbott:2009kk}. No significant excess was found, meaning that no coincident GW+HEN event was observed. 

The obtained nondetection can be used in an astrophysical context to place limits on the  density of joint GW+HEN emission events within the horizon of HEN and GW detection. The size of this horizon is related to both GW and HEN detection efficiencies, and can be estimated for typical emission models. In this particular analysis, two generic classes of models were considered: the final phase of a binary coalescence (short GRB-like, SGRB), or the collapse of a massive object (long GRB-like, LGRB), both followed by the emission of a relativistic hadronic jet responsible for the emission of HENs. To assess the GW sensitivity, simulated signals were injected into the on-source data. The injected signals consisted of inspiral waveforms (SGRB class) or Gaussian-modulated sinusoids at different frequencies (LGRB class). The amplitude of the minimal detectable signal can then be translated into an exclusion distance for the source. In both scenarios\footnote{Notice that for sine-Gaussian signals, the  optimistic assumption of a fixed energy of $E_{iso}^{GW} = 10^{-2} M_\odot c^2$ emitted in gravitational waves was used in that work.}, typical horizon distances are found to be in the range of 1 -- 20 Mpc. The HEN horizon of ANTARES for this search was estimated using the GRB flux predictions by~\citet{guetta04:_neutr_batse} and found to be of the same order of magnitude. From these estimates,  \citet{AdrianMartinez:2012tf} inferred limits on the population density of $\sim 10^{-2} $ Mpc$^{-3}$ yr$^{-1}$ for SGRB-like sources and $\sim 10^{-3} $ Mpc$^{-3}$ yr$^{-1}$ for LGRB-like sources, to be compared with current estimations of the local rate of potential progenitors, such as binary mergers for SGRB ($\sim 10^{-6}$ Mpc$^{-3}$ yr$^{-1}$, see~\citet{Kalogera:2003tn} and~\citet{Belczynski:2011qp}) or Type Ibc and Type II core-collapse supernovae for LGRB  (respectively $\sim 2\ 10^{-5}$ Mpc$^{-3}$ yr$^{-1}$, see~\citet{Guetta:2006gq} and $\sim 2\ 10^{-4}$ Mpc$^{-3}$ yr$^{-1}$, see~\citet{2009A&A...499..653B}).  

While the limits from this analysis using first-generation detectors still appear too high by a few orders of magnitude, one can expect that similar studies conducted with enhanced detectors will reach the required sensitivity to start constraining the astrophysical parameters, at least for some classes of models. Such an improvement will be made possible not only thanks to the increase in  sensitivity of the detectors themselves, but also thanks to the implementation of new strategies for the data analysis. Reconstruction strategies yielding a better angular accuracy for the HEN candidates, and GW search algorithms that can afford the computational cost of processing $\mathcal{O}(1O^3)$ candidate events, e.g., will be important ingredients for a jointly optimized GW+HEN search with ANTARES 12L, Virgo VSR2/3 and LIGO S6 data.  


  %

\subsection{Baseline search with combined HEN and GW events lists}
\label{subsec:parallel}

A baseline search with independent lists of HEN and GW candidates, as described in \citet{Baret:2011nu}, provides an advanced and interesting perspective for data analysis compared to more traditional externally triggered searches, e.g., with electromagnetic (EM) GRB observations. While EM observations of GRBs allow for searches for GW or HEN signals from a precisely determined time and direction, the joint search for GW and HEN signals with no EM counterpart relies on the combination of significance and directional probability distribution from these two messengers. Similar to how multiple GW detectors are effective in rejecting ``glitches" from the non-Gaussian background noise by requiring the coincident occurrence of an astrophysical signal in spatially separated GW detectors, requiring spatial and temporal coincidence from independent GW and HEN signal candidates can greatly reduce false alarm rate \cite{ligo_icecube}. The possibility that multiple neutrinos are detected from the same astrophysical source can also be considered, e.g. if several neutrino candidates happen to fall within a predefined space-time window.

Due to the uncertainty of directional reconstruction, especially for GWs, one can also enhance background rejection by using the expected source distribution in the nearby universe~\cite{Baret:2011nu}. Such distribution is not uniform; it can be based on the distribution of nearby galaxies and their weight. The density of at least some GW+HEN sources can be connected to the blue luminosity of galaxies \cite{blueluminositybinary1991ApJ...380L..17P,0264-9381-28-8-085016}, while source density can also depend on the type of the galaxy \cite{2006IJMPD..15..235D,2010ApJ...716..615O}. For galactic sources, the matter distribution within the Milky Way could also be taken into account. 

\begin{figure}
\begin{center}
\resizebox{0.6\textwidth}{!}{\includegraphics{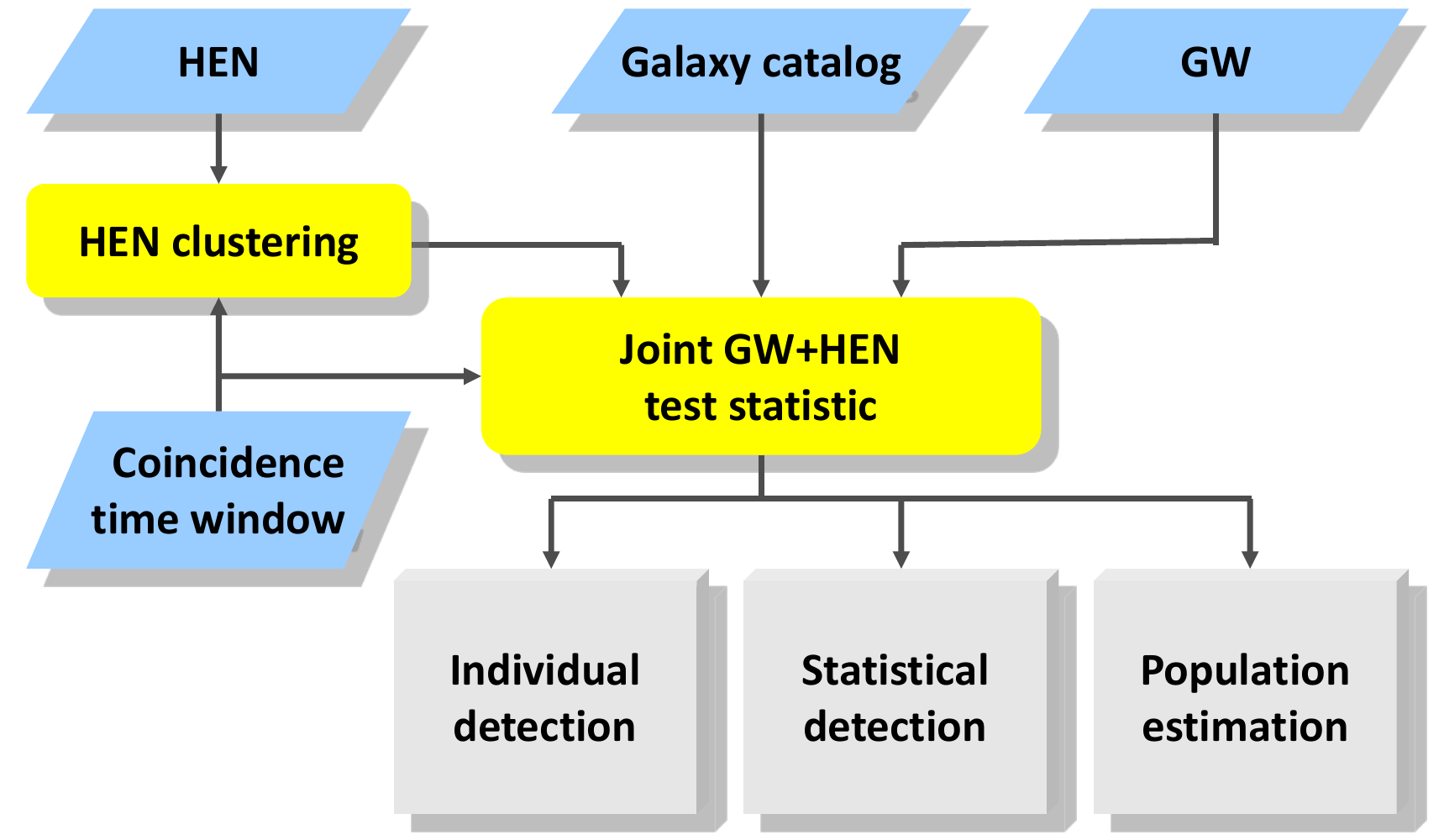}}
\end{center}
\caption{Schematic flow diagram of a joint GW+HEN search pipeline. The inputs of the pipeline are, besides data from HEN and GW detectors, the astrophysical source distribution from a galaxy catalog, as well as the coincidence time window used for the search. Spatially and temporally coincident neutrinos can be clustered, potentially increasing the significance and decreasing the false alarm rate of a coincident GW+HEN signal. Combining this information in a joint test statistic, one can evaluate the results to look for individual or statistical detection of signal candidates. Upon non-detection, the results can be used to determine an upper limit on the source population.} \label{figure:flowchart}
\end{figure}

\begin{figure}[!t]
\vspace*{-0.2cm}
\includegraphics[width=0.52\linewidth]{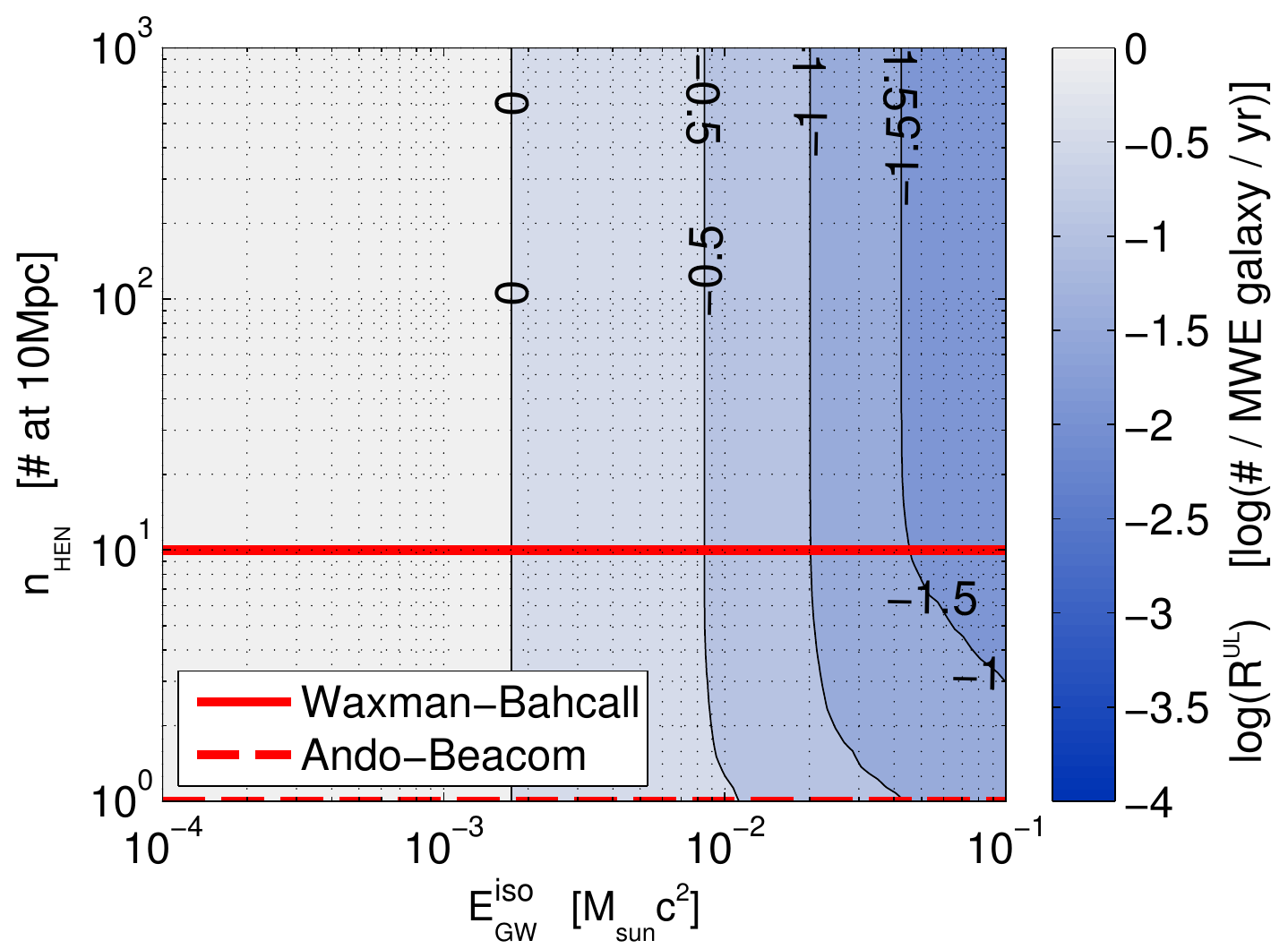}
\hfill
\includegraphics[width=0.47\linewidth, height=7cm]{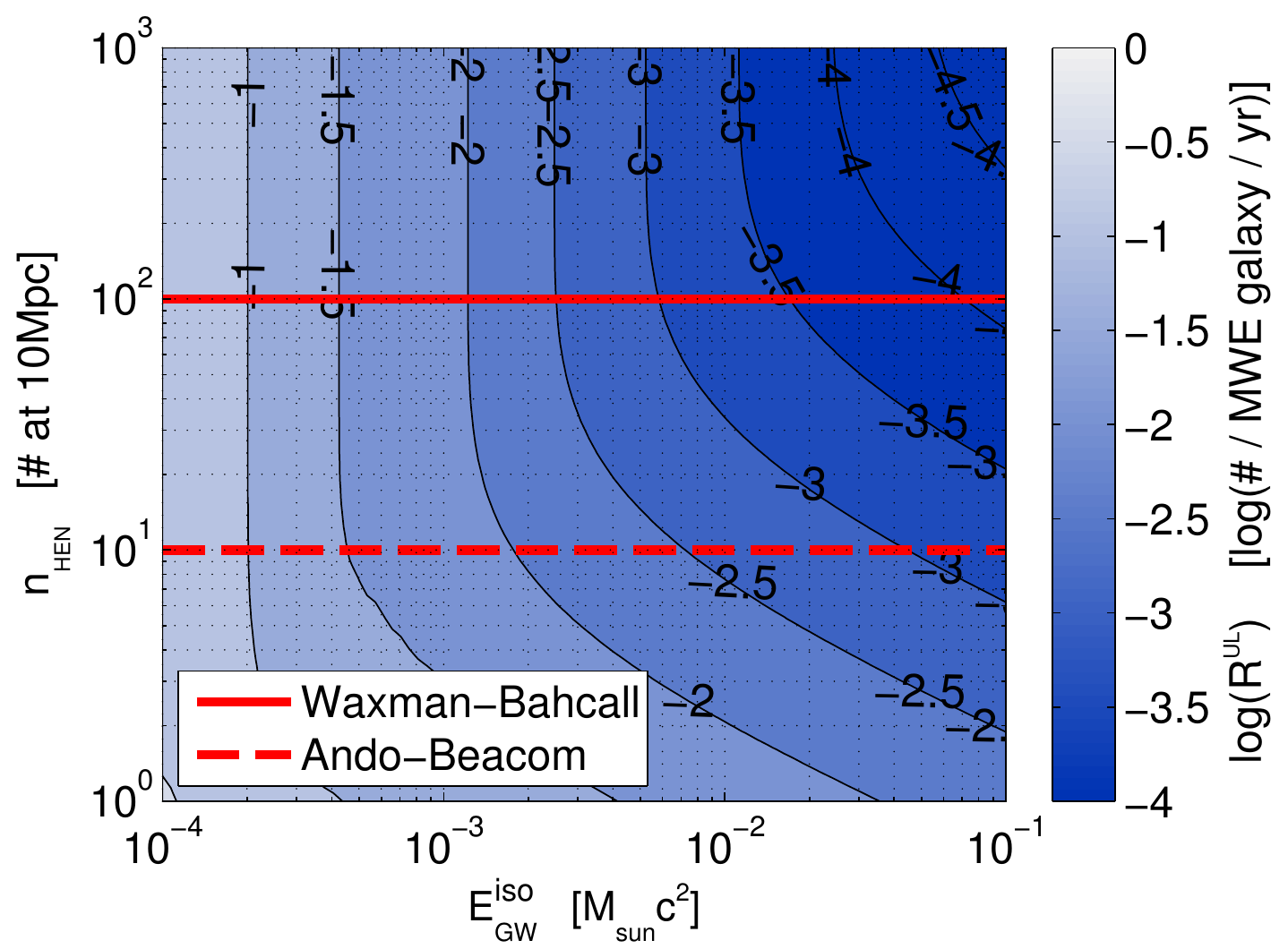}
\caption{Expected upper limits for a GW+HEN source population as inferred from a joint GW+HEN search with 1 year of coincident observation time, respectively, with IceCube IC22 and initial Virgo/LIGO detectors (left panel), and with full IceCube IC86 and advanced Virgo/LIGO detectors (right panel). The results take into account the blue-luminosity-weighted galaxy distribution. The benchmark source parameters are $E_{iso}^{GW}$ (total energy emitted in GW) and $n_{HEN}$ (number of detected neutrinos from a source at 10 Mpc). The population rate upper limit (color scale) is expressed in logarithmic units of number of sources per Milky Way-equivalent galaxy per year. The horizontal lines show predictions for $n_{HEN}$ from the emission models of~\citet{WBfireball} and~\citet{2005PhRvL..95f1103A}.
From~\citet{Baret:2011nu}.}
\label{fig:limpop}
\end{figure}

The search for joint GW+HEN signal candidates can aim for the detection of a single astrophysical signal with high enough significance to claim detection. Another possibility is to aim for a set of weaker signals that could not be detected individually, but which have a joint distribution that differentiates them from the background. Such a technique has been used in various searches for GWs; see e.g. \citet{Abbott:2009kk}. A schematic flow diagram of a joint search algorithm is shown in Fig.~\ref{figure:flowchart}; more detail can be found in~\citet{Baret:2011nu}. 
The outcome of such a search can also be used to constrain the population of astrophysical GW+HEN sources, as described in~\cite{2011arXiv1108.3001B}. As an illustration, Fig.~\ref{fig:limpop} shows the expected upper limits on the GW+HEN source population that could be inferred from 1 year of coincident observation time with initial and advanced detectors respectively. 


\section{Conclusions}

GWs and HENs both carry information from the depth of their astrophysical source that is, to a large extent, complementary to the information carried by electromagnetic radiation. While the GW signature of cosmic events is characteristic of the dynamics of their central engine, a HEN flux is reflective of the hadronic outflow driven by the central engine. Detecting both messengers from a common source would provide the unique opportunity to develop and fine tune our understanding of the connection between the central engine, its surroundings, and the nature of the outflows.

Astrophysical targets of GW and HEN searches include gamma-ray bursts, soft-gamma repeaters, supernovae, and other intriguing transients. 
Such sources are often expected to be observable through electromagnetic  messengers, such as gamma-rays, X-rays, optical and radio waves. Some of these channels have already been  used both in searches for GWs with the LIGO-GEO-Virgo interferometer network, and in searches for HEN with the current neutrino telescopes ANTARES and IceCube. However, many of the emission models for these astrophysical objects have so far been indistinguishable by electromagnetic observations. The combination of two weakly interacting messengers, GWs and HENs, therefore provides a new and exciting opportunity for multimessenger searches with different challenges and outcomes from observations using electromagnetic counterparts.  

Such a joint GW+HEN analysis program could significantly expand the scientific reach of both GW interferometers and HEN telescopes. The robust background rejection arising from the combination of two totally independent sets of data results in an increased sensitivity and the possible recovery of cosmic signals. The observation of coincident triggers would provide strong evidence for the existence of common sources, some of which may have remained unobserved so far by conventional photon astronomy. Information on the progenitor, such as trigger time, direction and expected frequency range, can also enhance our ability to identify GW signatures or astrophysical HENs with significances close to the noise floor of the detectors.


More generally, the joint (non-)observation of HEN and GW signals in current and future detectors can provide important clues that address the fundamental questions about the nature and behaviour of violent astrophysical sources~\cite{Marka:2010zz}: What are the engines that produce simultaneous high energy neutrino and gravitational wave emission? What is the population of GW+HEN emitters that are not detected via electromagnetic signals, and how (if at all) does this depend on GW or HEN emission? Are there any sub-populations? What are the earliest times of HEN emission compared to the onset of GW emission?  When possible, the information obtained from a GW+HEN search will be combined with electromagnetic observations in order to get a full, multimessenger picture of the astrophysical phenomena under investigation. Such an approach might for instance provide information on jet propagation in the vicinity of the central engine, therefore probing the progenitor structure~\cite{2012PhRvD..86h3007B}. Classifying the common progenitors of GWs, HENs and gamma rays is an exciting opportunity. For instance, if there are sub-populations of GRBs that are efficient neutrino emitters, it will be interesting to identify the central engines behind these sub-populations using GWs. Observations might also enable us to confirm that gamma-ray and HEN beaming are similar/identical. Upon detection, the relative times of arrival or relative flux of the different signals can indicate important properties of the central engine; on the other hand, the absence of coincident signals can constrain the joint parameter space of the source. In the promising case of GRBs, the outcome of a joint GW+HEN search could e.g. improve our understanding of the details of astrophysical processes connecting the gravitational collapse/merger of compact objects to black-hole formation as well as to the formation of fireballs.

Several periods of concurrent observations with GW and HEN detectors have already taken place and the corresponding data are being scrutinized for coincident GW+HEN signals. Future schedules involving next-generation detectors with a significantly increased sensitivity (such as KM3NeT and the advanced LIGO/Advanced Virgo  projects) are likely to coincide as well. Studies are ongoing to explore the reach of current and planned experiments in constraining the population of multimessenger sources of GWs and HENs, with or without electromagnetic counterpart \cite{ET,2011arXiv1108.3001B}. Constraints on the rate of GW and HEN transients can be derived using the independent observations already available from the current generation of detectors. On this basis, \citet{2011arXiv1108.3001B} estimated the reach of joint GW+HEN searches using advanced GW detectors and the completed IC86 detector, showing that searches undertaken by advanced detectors are indeed likely to be capable of detecting, constraining or excluding several existing astrophysical models within one year of observation.



\section{Acknowledgments}

 
 The authors are indebted to many of their colleagues for frequent and fruitful discussions. In particular, they are grateful to hundreds of collaborators for making these large scale experiments a reality. They also thank Alexa Staley, David Murphy and Maggie Tse for their careful reading of the last version of the manuscript.
 
 The authors gratefully acknowledge support from the Argentinian CONICET (grant PIP 0078/2010), Columbia University in the City of New York and the US National Science Foundation under cooperative agreement PHY-0847182, the European Union FP7 (European Research Council grant GRBs and Marie Curie European Reintegration Grant NEUTEL-APC 224898), the French Agence Nationale de la Recherche (ANR-08-JCJC-0061-01), the Israel-U.S.Binational Science Foundation, the Israel Science Foundation, the Joan and Robert Arnow Chair of Theoretical Astrophysics, the Spanish Ministerio de Educaci\'on y Ciencia (grant AYA2010-21782-C03-01), the Swedish Research Council and the UK Science and Technology Facilities Council (STFC), grants PP/F001096/1 and ST/I000887/1. 

This paper had been assigned the LIGO document
number P1100194.

\bibliographystyle{apsrmp4-1}
\bibliography{Ando}

\end{document}